\documentclass[nofootinbib,aps]{revtex4}

\usepackage{amstext,amsmath,amssymb,amsfonts}
\usepackage[dvips]{graphicx}
\usepackage{latexsym}
\usepackage{euscript}
\usepackage{epsfig}
\usepackage{psfrag}

\makeatletter
\@addtoreset{equation}{section}

\newcommand{\Ref}[1]{(\ref{#1})}
\def \subs{\subsection}

\newcommand{\N}{\mathbb{N}}

\newcommand{\C}{\mathbb{C}}

\DeclareMathOperator{\tr}{tr}

\newtheorem{prop}{Proposition}

\def\be{\begin{equation}}
\def\ee{\end{equation}}
\def\bes{\begin{eqnarray}}
\def\ees{\end{eqnarray}}
\def\nn{\nonumber}
\def\arr{\rightarrow}

\def\ra{\rangle}
\def\f{\frac}

\def\wt{\widetilde}

\newcommand{\SU}{\mathrm{SU}}
\newcommand{\U}{\mathrm{U}}
\newcommand{\SO}{\mathrm{SO}}

\def\hh{{\cal H}}

\def\rwm{{\rm RWM}}
\def\rw{{\rm RW}}

\def\6{\langle}
\def\9{\rangle}
\def\tr{{\rm tr}\,}
\def\half{\mbox{$\f 1 2$}{}}
\def\1{\mbox{1\hskip-.25em l}}
\newcommand{\lalg}[1]{\mathfrak{#1}}
\def\osp{\lalg{osp}}

\newcommand{\ket}[1]{|#1\rangle}

\def\ii{{\cal I}}
\def\d1{{}^1 d}
\def\dk{{}^s d}
\def\aa{{\cal A}}
\def\aaM{{\cal A}_{{\rm macro}}}
\def\aam{{\cal A}_{{\rm micro}}}

\begin{document}
\title{Quantum Black Holes: Entropy and Entanglement on the Horizon}
\author{{\bf Etera R. Livine}\footnote{elivine@perimeterinstitute.ca},
{\bf Daniel R. Terno}\footnote{dterno@perimeterinstitute.ca} }
\affiliation{Perimeter Institute, 31 Caroline St N, Waterloo, ON, Canada N2L 2Y5}

\begin{abstract}
\begin{center} {\small ABSTRACT} \end{center}
We are interested in black holes in Loop Quantum Gravity (LQG).
We study the simple model of static black holes: the horizon is made of a given number of identical elementary surfaces and these small surfaces all behaves as a spin-$s$ system accordingly to LQG.
 The chosen spin-$s$ defines the area unit or area resolution, which the observer uses to probe the space(time) geometry. For $s=1/2$, we are actually dealing with the qubit model, where the horizon is made of a certain number of qubits.
In this context, we compute the black hole entropy and show that the factor in front of the logarithmic correction to the entropy formula is independent of the unit $s$.
We also compute the entanglement between parts of the horizon. We show that these correlations between parts of the horizon are directly responsible for the asymptotic logarithmic corrections. This leads us to speculate on a relation between the evaporation process and the entanglement between a pair of qubits and the rest of the horizon. Finally, we introduce a concept of renormalisation of areas in LQG.

\end{abstract}
\maketitle

\tableofcontents

\newpage

\section{Introduction}

Loop Quantum Gravity (LQG) is a canonical quantization of General
Relativity, which relies on a 3+1 decomposition of space-time (for
reviews, check \cite{lqg}). It describes the states of 3d geometry
and their evolution in time (through the implementation of a
Hamiltonian constraint). The states of the canonical hypersurface
are the well-known {\it spin networks}, which represent polymeric
excitations of the gravitational field. The main and first
achievement of the LQG framework is the implementation of quantum
area operators and the derivation of their discrete spectrum. A
surface is now made of elementary patches of finite quantized
area. Based on this discrete structure of the 3d space, one can
study in details the entropy associated to a surface and the
entanglement between the patches making the surface. The main goal
of the present work is to study analytically the relation between entropy and
entanglement of the horizon state and discuss their physical/geometrical
interpretations.

\medskip
Let us discuss more precisely the concepts that were mentioned
above. {\it Entanglement} can be loosely defined as an exhibition
of stronger-than-classical correlations between the subsystems. For a long time, it stood up among the apparent ``paradoxes" of quantum
mechanics for its nonlocal connotations \cite{per}. Recently it
became one of the main resources of quantum information theory
\cite{per,nc}. We discuss its relevant properties in Sec. IIB.
In our work we are not concerned with its role in quantum
information processing. Instead we are interested in its
relations to the entropy of black holes  and the role of
entanglement as a pointer to their physical properties, similarly
to the recent discussion of quantum phase transitions \cite{qpt}.

A spin network is a graph, a network of points with links
representing the relations between points. Each link or edge is
labeled by a half-integer, which defines the area of a surface
intersecting the link. More generally, a surface is defined
through its intersections with the edges of the spin network
described the underlying quantum 3d space: the surface can be
thought as made from elementary patches, each corresponding to a
single intersection with a link and whose area is given by the
label of that link. These labels actually stand for $\SU(2)$
representations i.e {\it spins}, and are conventionally noted $j$.
Due to regularization ambiguities, there is not a definitive
consensus on the precise area associated to a given spin $j$. The
original and conventional prescription is of an elementary area
$a(j)\equiv \sqrt{j(j+1)}$, but other reasonable choices are
$a(j)\equiv j$ or $a(j)\equiv j+1/2$ (although one would like to
keep $a(j=0)=0$ in the end for both physical intuitive reasons and
also mathematical consistency \cite{corichi}). Now, points, or
vertices, of a spin network represents chunks of volume.
Mathematically, they are attached a $\SU(2)$ intertwiner - a
$\SU(2)$ invariant tensor between the representations attached to
all the edges linked to the considered vertex.

A generic surface on a spin network background is thus described
as a set of patches, each punctured by a unique link of the spin
network.  A $\SU(2)$ representation $j$ is attached to each patch.
We denote its Hilbert space $V^{j}$. Intuitively, a (quantum)
vector $|j m\ra$ in the Hilbert space corresponds to the
geometrical normal vector to the surface defined by the elementary
patch. Now the spin network defines how the patches, and therefore
the whole surface, is embedded in the surrounding 3d space and
describes how the surface folds. For a closed surface, the region
of the spin network which is inside the surface exactly defines an
intertwiner, invariant under $\SU(2)$, between the patches of the
surface. This intertwiner contains the rotation-invariant
information on the way the patches organize themselves to form the
surface e.g the angle between two patches.

An important remark is that any spin $j$ representation $V^j$ can
be decomposed as a  symmetrized tensor product of $2j$ spin-\half{}
representations $V^{1/2}$. Therefore, one can interpret that a
fundamental patch or elementary surface is a spin-\half{}
representation. All higher spin patches can be constructed from
such elementary patches. For example, considering two spin-\half{}
patches, they can form a spin 0 representation or a spin 1
representation: in one case, the two patches are folded on one
another and cancel each other, while in the later case they add
coherently to form a bigger patch of spin 1. Considering an
arbitrary surface, one can then look at it at the elementary
(fundamental) level decomposing it into spin-\half{} patches, or one
can look at it at a coarse-grained level decomposing the same
surface into bigger patches of spin $s> 1/2$. From this point of
view, the size of the patches used to study a surface is like the
choice of a ruler of fixed size used by the observer to analyze
the properties of the object. When studying the entropy and
entanglement on a given surface, we will therefore consider it
made of a certain number of patches $1/2$ at the elementary level
(and not allow the size of the patches to vary when performing
entropy computations). Then one can study the coarse-graining or
renormalisation of these quantities when one observes the surface
at a bigger scale, using bigger patches to characterize the
surface.

\medskip

In the present work, we would like to study some basic properties
of {\it Black Holes} in the framework of Loop Quantum Gravity.
Black holes are actually the main object studied in quantum
gravity. Indeed, on one hand, they are especially simple objects
from the point of view of theoretical general relativity, and on
the other hand we expect a consistent theory of quantum gravity to
shed light on the origin of the entropy of a black hole and on the
so-called ``information-loss paradox" \cite{wald,rmp}. Here we
propose to analyze the basic features attached to the horizon of a
black hole through a simple model inspired from Loop Quantum
Gravity.

We deal with the simplest case of a Schwarzschild (non-rotating)
black hole at the kinematical level. Actually, studying black
holes at the dynamical level in quantum gravity should be very
interesting, and should allow to understand the evaporation
process at the level of space-time. We will later comment on the
dynamics as induced by LQG, but this will not be the main point of
our study; instead we have decided to focus on the black hole
kinematics. Considering a static black hole, our main
assumption is that the only information accessible to the observer
outside the black hole can be seen on its (event) horizon.
Considering the horizon as a closed surface, LQG describes it as made of patches of quantized area and describes the
interior of the black hole in terms of (a superposition of) spin
networks whose boundary puncturing the horizon defines the patches
of the horizon surface. From the point of the external
observer, it does not matter what is inside the black hole,
but only the information which could be read off the
horizon is relevant. More mathematically, the horizon surface is defined as a
set of elementary patches to each of which is attached a given
$\SU(2)$ representation. Let us label the patches with an index
$i=1,..,n$ and note the corresponding spins $j_i$. The space
geometry within the horizon, i.e. the black hole, is fully
described by a spin network, which can possibly have support on a
complicated graph within the horizon. From the point of view of
the horizon, the details of the spin network inside the black hole
do not matter: what actually matters is which intertwiner between
the $\SU(2)$ representations on the horizon does the spin network
induce. Indeed the observer outside the black hole is only
interested by the state of the patches on the boundary, which
lives in the tensor product $V^{j_1}\otimes..\otimes V^{j_n}$. The
only constraint on the state is that it should be globally gauge
invariant, i.e. $\SU(2)$ invariant, such that the horizon state is
exactly described by an intertwiner between the representations
$V^{j_i}$. From this point of view, horizon states are described
by a spin network having support on a graph with a single vertex
inside the horizon, i.e. a fully coarse-grained spin network.
Actually, we would like to stress that our analysis applies to generic closed surface states as soon as we decide to ignore the details of the geometry of the enclosed spatial region.

The basic model for a quantum black hole that we propose to study
is not new. We assume that the model describe the black hole
horizon at the fundamental level. In other words, we assume that
the observer describing the black hole have the maximal possible
resolution and probes space(time) with a ruler of minimal area
i.e. corresponding to a spin 1/2 surface. The resulting model for
the horizon is to consider it made of a number $n$ of spin 1/2
patches: a horizon state will live in the tensor product
$(V^{1/2})^{\otimes n}$. More exactly, we want a $\SU(2)$
invariant state, i.e. an intertwiner between these $n$
representations. At this point, let us point out that a spin 1/2
system is usually called a {\it qubit} (in the language of quantum
information \cite{nc}), so that one can say that the horizon is
made of $n$ qubits and that a horizon state is a singlet state for
these $n$ qubits. Finally, the area of the horizon surface will by
definition be $n\times a_{1/2}$ where $a_{1/2}$ is the area
associated to an elementary spin $1/2$ patch.

In this model, we will present the calculation of the entropy of
the black hole and recover the entropy law, in an asymptotic
limit, with a first term proportional to the horizon area and a
logarithmic correction with a factor $-3/2$. For these purpose, we
develop a method to compute the number of intertwiners as a random
walk modified with a mirror at the origin. Then the point that we
would like to particularly stress is that the $n$ qubits on the
horizon are correlated, and more precisely entangled: the horizon
is {\it not} made of $n$ uncorrelated qubits and the $-3/2$ factor
is actually a reflection of the fact that these qubits are
entangled.

Then one can renormalize, or equivalently coarse-grain, this
fundamental model, now assuming that the observer probes the
space(time) with a spin $s\ge 1$ elementary surface.  A
coarse-grained model of the black hole is then to consider horizon
states to be given by intertwiners between $n$ representations of
a fixed spin $s$. One can repeat the entropy calculation and check
how the entropy law gets renormalized. The first term proportional
to the area should simply get scaled by the ratio of the area
$a_s$ corresponding to a spin $s$ surface to the $a_{1/2}$. Then
we will prove using the random walk analogy that the logarithmic
correction factor is universally $-3/2$ and does not get
renormalized.

\medskip

Our approach is different from the standard loop quantum gravity approach, which studies the classically induced boundary theory on the black hole horizon (or generically any isolated horizon). It is shown that one gets a $\U(1)$ Chern-Simons theory \cite{bhlqg}, which one can quantize and count the quantum degrees of freedom. In particular, some ways of computing the black hole entropy in such a context also involve parallels with random walk calculations \cite{jerzy}. Nevertheless, one would ultimately like to start with the full quantum gravity theory, identify horizons within the quantum states of geometry and study the quantum induced boundary theory on the identified quantum horizon. However, searching for horizons in a spin network states requires either introducing a well-defined notion of observer at the quantum level and/or being able to extract the (semi-classical) metric from the quantum state of geometry. We do not tackle these issues in the present work. We decide to model the black hole as a region of the quantum space with a well-defined boundary (the horizon) such that the only information about the geometry of the internal region accessible to the external observer is information which can be measured on the boundary. Our results can therefore be applied to any closed surface in loop quantum gravity. Issues about actually identifying this closed surface as a horizon at the semi-classical level is left for future investigation.

\medskip

The paper is organized as follows. The next section will detail
the fundamental spin-\half{} black hole model. We will compute the
entropy of the black hole and entanglement between qubits on the
horizon. This will lead us to relate the entanglement of a pair of
qubits (or more generally a small part of the horizon) with the
rest of the horizon with its probability to evaporate from the
black hole. We remark that the evaporation process can not work on
a single qubit so that minimally only pairs of spins $1/2$ can
evaporate. Also, introducing a time scale canonically associated
to a black hole, like the (minimal) time of flight of a particle
on a circular orbit around the black hole, we derive the rate of
evaporation from the probability computed previously. Section III
deals with the generalized spin $s$ models. Using the random walk
analogy, we derive the entropy law and show the universality of
the 3/2 factor for the logarithmic correction. Section IV tackles
the issue of the renormalisation of surface areas in the framework
of LQG. Consider a surface defined microscopically on a spin
network. The spin network describes how the surface is folded and
embedded in space. A natural issue  is what will be the
coarse-grained area that an observer will assign macroscopically
to that surface. We use the calculations of intertwiners to deduce
the most probable area seen macroscopically assuming no knowledge
of the underlying spin network. We then propose to refine this by
defining a (background independent) state of a surface in (loop)
quantum gravity through the probability amplitude assigned to
intertwiners between the elementary patches making the surface.
This information can be equally thought as data on the surrounding
spin network or as data on how the surface geometry gets
coarse-grained (how its geometry looks at different scales).


\section{The spin-1/2 black hole model} \label{spinhalf}

We describe the horizon of the black hole as a two-sphere
punctured by the underlying spin network by $2n$, $n\in\N$, edges
labeled by spins $\half$. The number of punctures defines the area
of the horizon: $\aa\equiv 2n a_{1/2}$, where $a_{1/2}$ is the
area corresponding to a spin-$\half$ elementary patch. The
geometry of the interior of the black hole  is described by the
potentially complex graph inside the two-sphere. However, this
information is not available to an observer living outside the
horizon. This external observer has only access to the information
on the horizon, i.e., to the horizon state. It is given by the
intertwiner between the $2n$ punctures on the horizon, which
corresponds to completely coarse-graining the internal spin
network into a single vertex. The state lives in
$(V^{1/2})^{\otimes 2n}$ and is $\SU(2)$-gauge invariant. The
tensor product of $2n$ spins $\half$ simply reads:
\be
\bigotimes^{2n}\,V^{1/2}
\,=\,
\bigoplus_{j=0}^n\,V^j\otimes \sigma_{n,j}, \label{schur1}
\ee
where $\sigma_{n,j}$ is the irreducible representation of the permutation group
 of $2n$ elements corresponding to the partition $[n+j,n-j]$. More details of the
  representation theory of the permutation group can be found in Appendix~D.
The dimensions $d_j^{(n)}\equiv {\rm dim}(\sigma_{n,j})$ count the degeneracy of each
spin $j$ in the tensor product $(V^{1/2})^{\otimes 2n}$. The intertwiner
space ${\cal H}$ is defined as the $\SU(2)$-invariant subspace of the tensor
 product i.e. the spin $j=0$ subspace. It is obviously isomorphic to $\sigma_{n,0}$.
  We note its dimension $N\equiv d_0^{(n)}$. Let us remind that we are considering an
  even number of spins since there does not exist any intertwiner between an odd number of $1/2$-spins.

It is known that the degeneracy factors are given in terms of the binomial coefficients:
\be
d_j^{(n)}=C_{2n}^{n+j}-C_{2n}^{n+j+1}=\f{2j+1}{n+j+1}C_{2n}^{n+j}.
\ee
In particular the dimension of the intertwiner space is
\be
N=\f{1}{n+1}C_{2n}^n.
\ee
Below we give a derivation of these formulas in terms of random walks,
 which will be straightforwardly generalizable to higher spin models
 when we replace the fundamental $1/2$-spin by an arbitrary $s$-spin.
Using the Stirling formula\footnote{We remind that Stirling formula gives the asymptotical behavior of $n!$:
$$
n!\sim\sqrt{2\pi n}\left(\f{n}{e}\right)^n.
$$} to compute the asymptotics of the Catalan number $C_{2n}^n/(n+1)$,
we obtain the asymptotic behavior of the entropy of the quantum black hole:
\be
S\,\equiv\,\ln N
\,\underset{n\arr\infty}\sim\,
(2n)\ln 2 -\f{3}{2}\ln(n)-\f{1}{2}\ln\pi. \label{statentr}
\ee
The first term is linear in $n$ and thus in the area, and is the
usual term that one expects to appear in any black hole entropy
calculation. The precise factor in front of the area $\aa$
actually depends on the unit used to define areas. The next term
is the logarithmic correction in $\ln(\aa)$. It is independent of
the unit used to measure areas and thus the $-3/2$ factor should be universal (invariant under coarse-graining). We will not discuss the other terms, which can be considered as negligible in the large area case. In the next
section, we will show that we get the same factor in any $s$-spin black hole model.

Under the assumption that the external observer has no knowledge of the internal geometry of
the black hole, the density matrix ascribed to the black hole horizon is the totally mixed state on the intertwiner space:
\be
\rho=\frac{1}{N}\sum_{i=1}^N|\ii_i\9\6\ii_i|,\label{rho}
\ee
where $|\ii_i\9$ form an arbitrary orthonormal basis of the
intertwiner space. It is obvious that the von Neumann entropy of
this state reproduces the (micro-canonical) entropy: $S(\rho)=-\tr\rho\ln\rho=\ln
N$. Below we study the entanglement properties of this state and
show how the logarithmic correction to the entropy formula
reflects the correlations between the $2n$ patches making the
black hole horizon.

Let us end this introduction by discussing the generality of this $1/2$-spin model.
 Thinking about rotating black holes, some might be tempted to only require invariance
 of the state under $\U(1)$ rotations around a particular axis -say $J_z$- and thus
 count all states with vanishing angular momentum $m=0$ instead of the stronger constraint
 of vanishing spin $j=0$. Then the dimension of the horizon state would be $C_{2n}^n$ and we
  would have only a $-1/2$ factor in front of the logarithmic correction. However the $\SU(2)$
  invariance which we require is the gauge invariance present in loop quantum gravity and has
  nothing to do the physical $\SO(3)$ isometry group of the two-sphere. A more serious criticism
   would be that in loop quantum gravity one must allow spin network labeled by arbitrary spins
    and thus we should more generally allow punctures of arbitrary spin and not restricting
     them to $1/2$-spins in order to have the true picture of a quantum black hole. Our
     counter-argument is that any spin $k$ can be decomposed into $1/2$-spins and thus
      allowing higher spin punctures would lead to an overcounting of the intertwiners
      and horizon states\footnotemark. More precisely, for example, an intertwiner between one
      1-spin and $(2n-2)$ $1/2$-spins automatically defines a unique intertwiner between
      $(2n)$ $1/2$-spins (through the unique intertwiner $V^{1/2}\otimes V^{1/2}\arr V^1$). From this
       perspective, one can consider the $1/2$-spin black hole model as the universal black hole
       description in loop quantum gravity. We will discuss this issue in more details in the
       later Sec.~\ref{renormalisation} on area renormalisation in loop quantum gravity.
       Let us nevertheless point out that the interested reader can find details
       on the structure of the intertwiner space when allowing arbitrary spins
       labeling the punctures in \cite{matrixmodel}. Another argument in favor
       of the $1/2$-spin black hole model is given in \cite{BHlee} as a symmetry
       argument in the semi-classical limit. This is related to the fact that the
       intertwiner space provides a representation of the permutation group, which
        can be interpreted as the discrete diffeomorphisms of the black hole horizon
        as made as $2n$ elementary patches. We will tackle this issue later in Sec.~III.2.

\footnotetext{One can argue that replacing two $1/2$-spins by one 1-spin actually
 modifies the area. This depends on the precise microscopic area spectrum. If one chooses the obvious linear spectrum $a_j\propto j$, then one avoids this issue. Nevertheless, the standard prescription in loop quantum gravity is $a_j\propto \sqrt{j(j+1)}$ which gives $$
\f{a_1}{2a_{1/2}}=\sqrt{\f{2}{3}}.
$$
If this case, the coarse-graining actually modifies the area. One should then keep in mind
that a rigorous entropy count should involve counting all the states of a given classical area ${\cal A}$
 up to an uncertainty $\delta {\cal A}$. The dependence of the chosen scaling of $\delta {\cal A}$ with ${\cal A}$ would then
 certainly change the details of the entropy law.}

\subsection{Entropy from random walk}

In this section, we show how one can compute the degeneracies of the tensor
 product decomposition and thus the entropy as a random walk calculation.
 Indeed, describing the decomposition of $(V^{1/2})^{\otimes 2n}$ iteratively,
 the analogy with the random walk is clear. If we are at the $k$-spin representation
 $V^{k}$, tensoring it with $V^{1/2}$, we obtain the representations of spin $k-1/2$
 and $k+1/2$. Only if we are at the trivial representation $k=0$,
 then $V^0\otimes V^{1/2}$ is simply $V^{1/2}$. This way, it appears
 that the degeneracy of the $k$-spin in the tensor product $(V^{1/2})^{\otimes 2n}$
 can be computed as the number of returns to the spot $k$ after $2n$ iterations of the
 random walk with a mirror (or a wall) at the origin 0. More precisely:

\begin{prop}
The degeneracy coefficients $d^{(n)}_j$ in the decomposition of
the tensor product $(V^{1/2})^{\otimes 2n}$ into the irreducible
spin-$j$ representations  equal the number of returns $\rwm(j)$ to
the spot $j$ after $2n$ iterations of the random walk of step
$1/2$ with a mirror in $j=0$ and starting at the origin. Moreover
the random walk with a mirror is given in terms of the standard
walk by the simple relation:
\be
\rwm_n(j)=\rw_n(j)-\rw_n(j+1),
\ee
where $\rw(j)$ is the number of returns to the spot $j$ after $2n$ iterations of the random walk.
\end{prop}
To prove this result, it is more convenient to deal with integer steps instead of switching between half-integers and
integers after each step. Then we actually consider the tensor product as $((V^{1/2})^{\otimes 2})^{\otimes n}$: at
 each step, we are tensoring with $(V^{1/2})^{\otimes 2}=V^0\oplus V^1$ i.e. moving either one step up or
 one step down or staying at the same spot. Then it is straightforward to check
 that $\rwm_n(\ge j)\equiv \sum_{k\ge j}\rwm_n(k)$ satisfies the same iteration law
 as the number of returns $\rw_n(j)$ of the standard walk without mirror at the origin.

Then as $\rw_n(j)=C_{2n}^{n+j}$, we recover the previous formula $d^{(n)}_j=C_{2n}^{n+j}-C_{2n}^{n+j+1}$.
 In particular, this shows the simple identity:
\be
\sum_j d^{(n)}_j\,=\,\rw_n(0)=C_{2n}^{n}.
\ee
Having these exact formulas, it is straightforward to derive the
asymptotics of $\rw_n(j=0)$ and $\rwm_n(j=0)$ when $n$ grows
large, and the result is $2^{2n}/\sqrt{n}$ and $2^{2n}/n\sqrt{n}$,
respectively.

On the other hand, it is possible to directly extract these asymptotics without computing the exact expression.
Indeed, working on the unit circle $\U(1)$ and using the orthogonality of the modes $e^{in\theta}$, one has:
\be
\rw_n(j)=\f{1}{2\pi}\int_{-\pi}^{+\pi}d\theta\,
e^{-i(2j)\theta}(e^{i\theta}+e^{-i\theta})^{2n}
=2^{2n}\times
\f{1}{2\pi}\int_{-\pi}^{+\pi}d\theta\,
e^{-i(2j)\theta}(\cos\theta)^{2n}.
\ee
For $j=0$, we evaluate this integral in the asymptotic limit $n\arr\infty$ using the saddle point approximation.
 First, we write
\be
\rw_n(0)
=2^{2n}\times\f{1}{\pi}\int_{-\f{\pi}{2}}^{+\f{\pi}{2}}d\theta\, (\cos\theta)^{2n}
=2^{2n}\times\f{1}{\pi}\int_{-\f{\pi}{2}}^{+\f{\pi}{2}}d\theta\, e^{2n\log(\cos\theta)}.
\ee
Then the exponent $\phi(\theta)=\log(\cos\theta)$   is always negative and has a unique fixed point
(which is actually a maximum) at $\theta=0$ with the second derivative $\phi^{(2)}(\theta=0)=-1$.
Therefore we can approximate the integral by:
\be
\rw_n(0)\underset{n\arr\infty}{\propto}
2^{2n}\f{1}{\pi}\int_{-\f{\pi}{2}}^{+\f{\pi}{2}}d\theta\, e^{-n\theta^2}
\sim \f{2^{2n}}{\sqrt{n}}\times \f{1}{\pi}\int_{-\infty}^{+\infty}dx\, e^{-x^2},
\ee
and derive the right asymptotic behavior. Following the same line of thought,
starting from
\be
\rwm_n(j=0)=\rw_n(0)-\rw_n(1)=
2^{2n}\f{1}{2\pi}\int_{-\pi}^{+\pi}d\theta\,
(1-e^{-i2\theta})(\cos\theta)^{2n},
\ee
and using that this expression is real, we get:
$$
\rwm_n(0)
=2^{2n}\f{1}{\pi}\int_{-\f\pi 2}^{+\f\pi 2}d\theta\, (1-\cos 2\theta)(\cos\theta)^{2n}
=2^{2n}\f{1}{\pi}\int_{-\f\pi 2}^{+\f\pi 2}d\theta\, (1-\cos 2\theta)e^{2n\log(\cos\theta)}.
$$
Now we have an integral of the type $\int d\theta f(\theta) e^{-n\phi(\theta)}$. The exponent $\phi(\theta)$ has only one fixed point at $\theta=0$. Then since $f(\theta)$ vanishes at $\theta=0$, the leading order in the asymptotics is given by the first non-vanishing derivative of $f$:
\be
\rwm_n(0)\underset{n\arr\infty}{\propto}
2^{2n}\f{2}{\pi}\int_{-\f\pi 2}^{+\f\pi 2}d\theta\, \theta^2 e^{-n\theta^2}
\sim
\f{2^{2n}}{n^{3/2}}\f{2}{\pi}\int_{-\infty}^{+\infty}dx\, x^2 e^{-x^2}.
\ee
This shows the $2^{2n}/n^{3/2}$ asymptotic behavior of the dimension of the intertwiner space and thus of the horizon state space. The same analysis will be generalized in the next section to any $s$-spin horizon model. The factor $-3/2$ of the logarithmic correction to the area-entropy law will appear to be universal and, as shown above, has a  simple explanation in terms of random walks.

Although the asymptotics in $1/\sqrt{n}$ and $1/n\sqrt{n}$ have a
natural interpretation from  the random walk perspective, a
simpler way to obtain integral formulas for the degeneracies is to
use the orthogonality of the $\SU(2)$ characters.
Indeed, we have:
\be
d^{(n)}_j=\int dg\,\chi_j(g)\left(\chi_{\f12}(g)\right)^{2n}
=\f{2}{\pi}\int_0^\pi \sin^2\theta d\theta\, \chi_j(\theta)\left(\chi_{\f12}(\theta)\right)^{2n},
\ee
where $dg$ is the normalized Haar measure on $\SU(2)$. The character $\chi_j(g)$ is the trace of the group element $g$ in the $j$-spin representation and is a  function of only the rotation angle $\theta$:
$$
\chi_j(\theta)=\f{\sin(2j+1)\theta}{\sin\theta}\, ,
\qquad
\chi_{\f12}(\theta)=2\cos\theta.
$$

\subsection{Entanglement} \label{Sent}
The black hole model we analyze allows to deal with  entanglement
in a finite-dimensional setting.  In addition, we consider only a
bipartite entanglement, i.e., entanglement between two
distinguishable parts of this system. This is the best
understood part of the entanglement theory \cite{nc,ent}. In this
section and in the appendices we present only the most essential
elements of the analysis.
The missing proofs and the generalizations to other families of states
can be found in \cite{lit1}.

Pure state entanglement is easily identified. Pure states that
cannot be written as a direct product of two states are entangled.
For example, the singlet state of two qubits,
\be
|\Psi^-\9=\frac{1}{\sqrt{2}}\left(|+-\9-|-+\9\right),\label{singlet}
\ee
is an entangled state. The definition is more involved for mixed
states. It can be expressed as a part of the threefold hierarchy.
At the lowest level there are direct product states  $\rho=\rho_1\otimes\rho_2$.
Then, there are classically correlated (or separable) states,
\be
\rho=\sum_i w_i\rho^i_1\otimes \rho^i_2, \qquad \sum_i w_i=1,
\qquad \forall w_i>0.
\ee
which are mixtures (convex combinations) of the direct product states.
Finally, entangled states are those can not be represented as separable. They have their own hierarchy which does
not concern us here. Apart from two-qubit systems, no
single universal criterion  that  allows  to tell whether a mixed
state is entangled or not is known.

For pure states there is a natural way to quantify entanglement.
The degree of entanglement of a pure state $|\Psi\9$ is the von
Neumann entropy of either of its reduced density matrices,
\be
E(|\Psi\9\6 \Psi|)=S(\rho_{A,B})=-\tr \rho_{A,B}\log\rho_{A,B},
\qquad
\rho_A=\tr_B|\Psi\9\6 \Psi|.
\ee
For example, reduced density matrices of $|\Psi^-\9$ are the maximally
mixed spin-$\frac{1}{2}$ states $\rho_{A,B}=\1_{2\times 2}/2$, so
its degree of entanglement is $\log2=1$ bit. In general, a pure
state whose degree of entanglement is $\log d$, where $d$ is the
dimension of the  Hilbert space of one of the subsystems, is
called maximally entangled.

There are various measures of entanglement of mixed states that
reflect different aspects of their preparation. We adapt here the
{\em entanglement of formation}, which is defined as follows. For
all possible decompositions of the state $\rho_{AB}\equiv\rho$
into mixtures of pure states,
\be
\rho=\sum_i w_i |\Psi_i\9\6\Psi_i|, \qquad \sum_i w_i=1,
\qquad \forall w_i>0,
\ee
 the weighted average of degrees of entanglement of the
 constituents is calculated, and the minimum is taken over all
 decompositions
\be
E_F(\rho)=\min_{ \{ \Psi_i\}}\sum_i w_i S(\rho_i).
\ee
For pure states this expression reduces to the degree of
entanglement. Entanglement of formation is zero for unentangled
states. For generic states, an analytic expression for $E_F(\rho)$ exists only for  two-qubit systems.

Nevertheless, we were able  to calculate entanglement  of $\rho$
of Eq.~(\ref{rho})  for all bipartite splittings (Proposition 2
below). Moreover, it is possible to show that all standard
entanglement measures coincide on $\rho$ \cite{lit1}, so there is
no arbitrariness the choice of the entanglement measure.
Henceforth we drop the subscript and simply write $E(\rho)$.

Let us split the $2n$ qubits into a group of $2k$ and a group of
$2n-2k$ qubits and adapt the angular momentum basis. Hence the
Hilbert space is $(V^{1/2})^{\otimes {2k}}\otimes
(V^{1/2})^{\otimes {(2n-2k)}}$, for which a shorthand
$\hh=\hh_A\otimes\hh_B$ will be used. The basis states in either
of the subspaces are labeled as $|j,m, a_j\9$ and $|j,m, b_j\9$,
respectively. Here $0\leq j\leq k$ and $-j\leq m\leq j$ have their
usual meaning and $a_j$ enumerate different irreducible subspaces
$V^j$. Their multiplicities $d^{A,B}_j$ are given in Appendix B
and $a_j=1,\ldots, d^A_j$ and $b_j=1,\ldots d^B_j$. The constraint
${\bf J}^2=0$ ensures  the states $|\ii_i\9$ are the singlets on
$V^j_A\otimes V^j_B$ subspaces for $j=0,..,k,$
\be
|j,a_j,b_j\9\equiv\frac{1}{\sqrt{2j+1}}\sum_{m=-j}^j
(-1)^{j-m}|j,-m,a_j\9\otimes|j,m,b_j\9. \label{not1}
\ee
 A more transparent notation is based on representing each subspace
 with a fixed $j$ as $V^j\otimes D^j$, where $V^j$ is a $2j+1$
 dimensional space that carries a spin-$j$ irreducible representation (irrep) of SU(2) and
 $D^j$ is the degeneracy subspace. In the case of qubits the Shur's
 duality~\cite{good}
 identifies these spaces as
 irreps of the permutation group $S_{2n}$ that correspond to the partition $[n+j,n-j]$ of $2n$
objects \cite{lit1, wkt}. Hence each basis state can be
 represented as a tensor product state on three subspaces,
\be
|j,a_j,b_j\9\equiv|j\9_{AB}\otimes|a_j\9_{D^j_A}\otimes|b_j\9_{D^j_B}
\ee

 In this notation
Eq.~(\ref{rho}) becomes
\be
\rho=\frac{1}{N}\sum_j \sum_{a_j,b_j} |j,a_j,b_j\9\6 j,a_j,b_j|=
\frac{1}{N}\sum_j \sum_{a_j,b_j}|j\9\6
j|_{AB}\otimes|a_j\9\6 a_j|_{D^A}\otimes|b_j\9\6 b_j|_{D^B}
\ee
Reduced density matrices of these states,
 \be
\rho_{j,a_j}=\tr_B |j,a_j,b_j\9\6 j,a_j,b_j|,
 \ee
 where here and in the following the tracing out is over
$\hh_B$, are independent of $b_j$. There are exactly $d^B_j$ of
the matrices $\rho_{j,a_j}$, so the reduced density matrix
$\rho_A=\tr_B\rho$ is given by
\be
\rho_A=\frac{1}{N}\sum_j \sum_{a_j}d_j^B\rho_{j,a_j}.
\ee
 With this notation, it is possible to prove the following result:
\begin{prop}
\label{BHentprop}
The entanglement  of the black hole state $\rho$ is given by
\be
E(\rho)=\frac{1}{N}\sum_j \sum_{a_j}d^B_j
S(\rho_{j,a_j})=\frac{1}{N}\sum_j d_j^Bd_j^A\log(2j+1).
\ee
\end{prop}
The proof  can be found in Appendix A. Following \cite{lit1}, this
result can actually be generalized to more general zero-spin states:
\begin{prop}
\label{entG}
Consider an arbitrary convex combination of the zero spin states,
\bes
\rho= & 
\sum_j \sum_{a_j,b_j}w^{(j)}_{a_j,b_j}|j\9\6
j|_{AB}\otimes|a_j\9\6 a_j|_{D^A}\otimes|b_j\9\6 b_j|_{D^B},\nn
\\
 & \sum_{a_j,b_j}w^{(j)}_{a_j,b_j}=1
\ees
all measures of entanglement for the state $\rho$ are equal to
\be
E(\rho)=\sum_{j,a_j,b_j} w^{(j)}_{a_j,b_j}\log(2j+1).
\ee
\end{prop}

As an example we  consider the entanglement between the block of 2 qubits and the rest. Dividing the system into $2$
and $2n-2$ qubits leads to the following degeneracies:
\be
d_{0,1}^A=1, \qquad d_0^B={2n-2\choose n-1}\frac{1}{n}, \qquad
d_1^B={2n\choose n}\frac{3(n-1)}{2(2n-1)(n+1)}, \label{s02f}
\ee
that asymptotically satisfy $d_0^B/N\sim1/4$, $d_1^B/N\sim 3/4$.
Hence
\be
E(\rho |2)=\frac{d_1^B}{N}\log3\sim\frac{3}{4}\log 3.
\ee
Entanglement of 4 and $2n-4$ qubits is hardly more challenging,
with asymptotic values of the degeneracies being
\be
s^{(4)}_0\equiv d^A_0 d^B_0/N\sim1/8, \qquad s^{(4)}_1\equiv d^A_1
d^B_1/N\sim 9/16, \qquad s^{(4)}_2\equiv d^A_2 d^B_2/N\sim 5/16.
\ee
As a result
\be
E(\rho |4)\sim \frac{9}{16}\log3+\frac{5}{16}\log 5.
\ee
Fig.~1 illustrates the convergence of the entanglement to its
asymptotic value.
\begin{figure}[htbp]
\epsfxsize=0.45\textwidth
\centerline{\epsffile{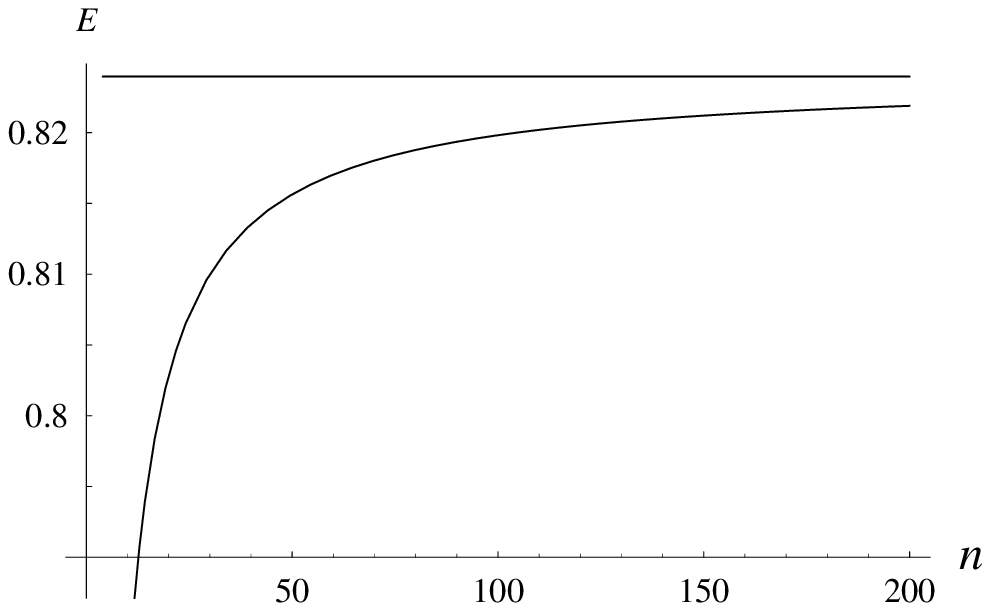}\epsfxsize=0.47\textwidth\epsffile{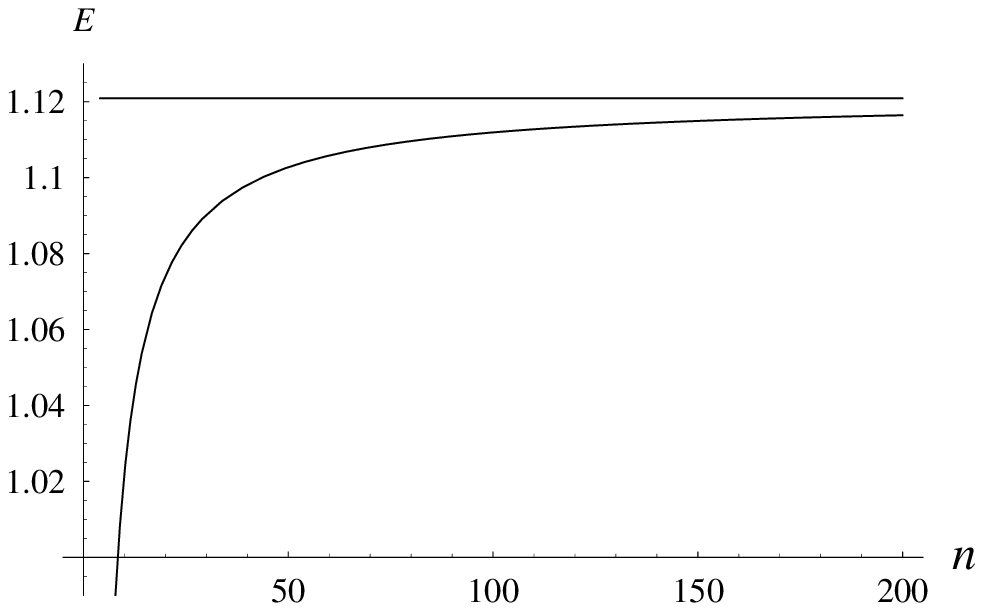}} 
\caption{\small{Entanglement between $2k$ and $2n-2k$ qubits for $2k=2$ (left) and $2k=4$ (right).  The entanglement of the two qubit block for $n=4$ is 0.706 ebits, and the entanglement of the four qubit block for $n=6$ is 0.9429 ebits.}}
\end{figure}

One finds that if the $2n$ qubits of the horizon are separated
into two sets of $n$ qubits each, then the entanglement of the state
$\rho$ (asymptotically) equals to $E(\rho|n)\sim \half
\log n$, while the  entropy $S(\rho)$  the reduced density matrices $\rho_n$ of its halves
is $S(\rho_n)\sim n\log 2$. More technically, this translates to the {\em quantum mutual information} \cite{ca} between the black hole and its halves being three times the entanglement between the
two halves:
\be
I_\rho(n:n)\equiv S(\rho_n)+S(\rho_n)-S(\rho)\sim 3E(\rho|n).
\label{qmi}
\ee
Let us  recall that for any pure entangled state $|\Psi\9$, like the singlet state of Eq.~(\ref{singlet}), we have:
\be
I_\Psi(A:B)\equiv 2E(\Psi).
\ee
In a mixed entangled state the classical correlations are present
on top of the entanglement, so it is natural to expect that
$I_\rho(A:B)>2E(\rho|A:B)$. From the quantum
information-theoretical point of view, it can be argued that the
quantum mutual information represents the total amount of
classical and quantum correlations \cite{gpw}. Hence our result
shows that the logarithmic correction to the black hole
entropy represents the total amount of correlations for the
symmetric splitting of the horizon state.

More generally, while the exact entanglement
between the two parts of the spin network depends on $k$, an explicit
calculation shows that Eq.~(\ref{qmi}) can be extended to
\be
I_\rho(A:B)\approx 3E(\rho|A:B) \label{qmi3}
\ee
for any nearly arbitrary partitions into two parts, $A$ and $B$.
For example, for $k=10$ the coefficient is approximately $2.956$.

\subs{About evaporation}

Let us now focus on the possible physical meaning of the
entanglement calculations computed above. Let us consider a pair
of qubit. It is in  a mixture of having an intertwiner $j=0$ or
$j=1$ with the rest of the horizon patches. In the case that it
has an intertwiner $j=0$, standard loop quantum gravity tells us
that the pair of qubits is actually not linked -{\it detached}-
from the rest of the horizon qubits i.e. it is {\it not} on the
horizon anymore. From the information point of view, a $j=0$
intertwiner means that the pair of qubits is uncorrelated with the
rest of the horizon, so intuitively it can not be part of the
black hole horizon. The entanglement, or more precisely the
entangled fraction, is a precise quantification
 of how much the pair of qubits is in the intertwiner state $j=1$ compared to being in in the intertwiner state $j=0$. It is therefore natural to assume a relation between the unentangled fraction and the evaporation probability of the pair of qubits (more precisely that there exists a monotonous function relating the two).

Let us consider the unentangled fraction of a pair of qubits $s_0^{(2)}$. It depends
of course on the total number of patches $n$, i.e. on the area of the black hole horizon.

The probability of evaporation of the pair of qubit is given by the {\it Born rule}:
\be
{\rm Prob}_{\rm evap}\,=\,
{\rm Tr}\,(\rho P_{j=0}),
\ee
where $\rho$ is the black hole density matrix and $P_{j=0}$ the projector on the subspace $\hh_{j=0}$ of intertwiners having a link of spin $j=0$ between the chosen pair of qubits and the rest of the horizon. Then $\rho$ is actually the identity on the intertwiner space $\hh$ normalized by the dimension $N=\dim\hh$, so that the probability is exactly computed as the unentangled fraction:
\be
{\rm Prob}_{\rm evap}\,=\,
\f{\dim\hh_{j=0}}{\dim\hh}=\f{d_0^{A={\rm 2 qubits}}d_0^{B={\rm rest}}}{N}=s_0^{(2)}.
\ee
Then to go from probabilities to
evaporation rate we need a time scale. In our simple model, we do not have any dynamics. The only way to
introduce a time scale is to go to the semi-classical limit: for large black holes, the natural time scale
is proportional to its mass $\tau\propto M \propto \sqrt{\aa} \propto\sqrt{n}$. For example, we can consider
the black hole as a clock with its period being the time of flight of the shortest stable circular trajectory
around the black hole. Finally, we can conclude that:
\be
\f{{\rm d}\aa}{{\rm d} t}\propto
\f{{\rm d}n}{{\rm d} t}\propto
-\f{s_0^{(2)}}{\tau}\propto
-\f{s_0^{(2)}}{M}\propto
-\f{s_0^{(2)}}{\sqrt{n}},
\ee
where the proportionality coefficients actually depend on the
precise choice of time scale and the exact relation between $n$,
the area $A$ and the mass $M$. Since $s_0^{(2)}$ converges to a
fixed non-zero value $1/4$ when $n$ grows to infinity, we get a (approximately) fixed
probability of evaporation (of the pair of qubits) for large black
holes. To make the analysis more exact, we should consider the
exact dependence of the unentangled fraction $s_0^{(2)}$ on the
black hole size $n$. Actually Eq.~(\ref{s02f}) gives us:
\be
s_0^{(2)}\sim\frac{1}{4}+\frac{3}{8n}.
\ee
which leads to a correction to the usual evaporation formula:
$$
\f{{\rm d}A}{{\rm d} t}\propto
\f{{\rm d}n}{{\rm d} t}\propto
-\f{1}{4\sqrt{n}}-\f{3}{8n\sqrt{n}},
$$
This might allow us to derive precise corrections to the
evaporation formula of black holes. Nevertheless, we should then
also consider the corrections to the semi-classical mass formula
$M\propto \sqrt{n}$. Moreover all relies on the use of a
semi-classical time scale, which would need to be derived from the
exact dynamics of the underlying quantum gravity theory. At the
end of the day, we recover in our simple model the standard
Hawking evaporation formula:
$$
\f{{\rm d}M}{{\rm d} t}\propto -\f{1}{M^2}.
$$
Our analysis is actually very similar to the one performed by Bekenstein and Mukhanov \cite{bm}.

We would like to point out that we have made the implicit assumption that the evaporation process is completely dominated by the evaporation of a pair of qubits. Indeed, to be thorough, we should consider as well the possible evaporation of bigger blocks. Nevertheless, we have computed the unentangled fraction of a block of $2k$ qubits (we only consider blocks of even size because of the $\SU(2)$ invariance requirement):
$$
s_0^{(2k)}\sim\f{1}{\sqrt{\pi}}\left(\f{n}{k(n-k)}\right)^{\f32}.
$$
And it is clear that the unentangled fraction drops rapidly with
the block size $k$. Actually, only one out of $2^{2k}$ basis
states of the $2k$-qubit block does not contain smaller
unentangled pieces, hence the ``proper" unentangled fraction (probability of a block of size $k$ to get detached without any of its smaller piece getting detached itself) is $2^{-2k}s_0^{(2k)}$, which now drops exponentially with the block size $k$.

We would like to insist on the fact that our analysis is not dynamical: we rely on the Born rule and treat the black hole evaporation as a traditional decay (like for radioactive emission).
 To go further, one should consider the geometry state of the interior
 and exterior of the quantum black hole, and not only the horizon state, and analyze
  their quantum dynamics. Our main point here is that it does make sense to relate the notion of entanglement
   between parts of the horizon to probabilities of evaporation from the horizon.

A last point is that the pair of qubits which evaporates will likely not be the Hawking radiation: it forms
a part of spacetime which is in the exterior of the black hole.
The Hawking radiation will be produced by the difference of energy
 between the system of a black hole with horizon size $2n$ and the
 system of a black hole with horizon size $2(n-1)$ plus the detached pair of qubits.

Finally, we have related the entanglement calculations between parts of the horizon to the Hawking evaporation of black holes. As entanglement is a purely quantum notion, this fits with the fact that the evaporation is a purely quantum effect. Moreover, it points toward a link between (quantum) information concepts such as correlation and entanglement and the physics of (quantum) geometry.

\subsection{Black holes at the interplay between  quantum gravity and quantum information}

As we have illustrated above, quantum information concepts, such
as  entanglement, are useful for the study of black holes in
quantum gravity. This relation  goes beyond conceptual
points and borrowing the technical tools of the entanglement
theory \cite{lit1}. In particular, the use of $\SU(2)$ invariant
subspaces in quantum black hole has a precise analogy in the
establishing of quantum communication without a shared reference
frame \cite{stevie}.

 Communication without a shared reference frame (SRF)
can be formulated as follows. Assuming that there are no sources
of noise, different orientations  of the reference frames of the
communicating parties induce a unitary map between their states
and operators that describe their experimental procedures. This
unitary representation depends on the physical nature of the
qubits, usually massive spin-$1/2$ particles. When the parties
have no knowledge about the respective orientation of their
frames, the state of the transmitted qubits, the initial state
$\rho$ looks averaged over $\SU(2)$ as far as the receiver is
concerned:
\be
\Upsilon(\rho)
=\int_{\SU(2)} dU\, U\rho U^\dag,
\ee
where $dU$ is the Haar measure of $\SU(2)$.
The task is to transfer information in the most efficient way despite this decoherence.

Indeed, the transmittable information   is to be encoded in a
$\SU(2)$-invariant way. For this purpose, to start with, we can
use the $j=0$ subspaces of the space of the $n$ transmitted
qubits. For example, in the decomposition of $(V^{1/2})^{\otimes
4}$, there are two distinct $j=0$ subspaces, which can be be used
to encode the states $\ket{+}$ and $\ket{-}$ of a single qubit. In
the notation of Eq.~(\ref{not1}) it is given, e.g., by
\be
|+\9\rightarrow|j=0,1,1\9,\qquad |-\9\rightarrow|j=1,1,1\9.
\ee
As we have seen earlier, the Schur's duality~\cite{good} gives in general:
\be
\hh_n\equiv
\bigotimes^{n}\C^2
=\bigoplus_{j=0(\half)}^{n/2} V^j\otimes \sigma_{n,j}.
\ee
A $\SU(2)$ transformation on $\hh_n$ will act  on the spaces $V^j$
while leaving the degeneracy spaces $\sigma_{n,j}$ invariant. It
is then straightforward to check that the map $\Upsilon$ acting on
density matrices on $\hh_n$  will randomizes the $V^j$ sector
(mapping to the maximally mixed state) while preserving the
information stored in $\sigma_{n,j}$. More precisely, it is
possible to show the following:
\begin{prop}
Consider the space $\hh_n$ and the map $\Upsilon$ acting on the
space of density matrices over $\hh_n$. Let us consider the space
$\hh_{2n}\equiv \hh_n^{(A)}\otimes \hh_n^{(B)}$ and its
$\SU(2)$-invariant subspace $\hh^0_{2n}\equiv \sigma_{2n,j=0}$.
Then ${\rm Im}(\Upsilon)$ is given exactly by the reduced density
matrices defined as the partial traces over $B$ of density
matrices on $\hh^0_{2n}$:
$$
{\rm Tr}_B\left({\rm Den}(\hh^0_{2n})\right)={\rm Im}(\Upsilon_n).
$$
In particular, for $\rho\in{\rm Im}(\Upsilon_n)$, there exists a $\wt{\rho}$ density matrix on $\hh^0_{2n}$ such that $\rho={\rm Tr}_B(\wt{\rho})$. Then by definition, the entropy of $\rho$ is $S(\rho)=E(\wt{\rho}|A:B)$ the entanglement of the state $\wt{\rho}$.
\end{prop}

The most efficient transmission method is a block encoding, when
$L$ logical qubits are encoded into $n$ physical ones. To this end
the qubits are encoded into the degeneracy labels of the highest
multiplicity spin-$j_{\max}$ representation of $\SU(2)$. As shown
in Sec. 4, the highest multiplicity spin is asymptotically
$j_{\max}=\sqrt{n}/2$. The number of logical qubits is the
logarithm $\log_2$ of the dimension of the degeneracy space
corresponding to $j_{max}$, hence  the efficiency $L/n$
asymptotically approaches 1 as $1-n^{-1}\log_2 n$.

The same setting can be applied to cryptographic purposes \cite{crypt}. Imagine that Soraya tries to communicate to Marc and that they share a common reference frame to measure and send qubits. On the other hand, we consider an eavesdropper Tamara who does not have any information about that common reference frame. In this context, for all state $\rho$ that Soraya sends, Tamara will see $\Upsilon(\rho)$. One can define the quantum capacity of the secured quantum channel following \cite{crypt}. We are looking for the biggest possible Hilbert space $\hh\subseteq\hh_n$ such that for all density matrix $\rho$ with support in $\hh$, $\Upsilon(\rho)$ is the maximally mixed state $\rho_0$ on $\hh$ and Tamara will not be able to extract any information about the message which Soraya wants to send to Marc. The quantum capacity is then, as usual, the logarithm of the dimension of the maximal such Hilbert space $\hh$.

This is easily done using the tools introduced previously. If $\rho_0$ is the maximally mixed state on $\hh$ then $\log({\rm dim}\hh)=S(\rho_0)$. Moreover, there exists a state $\wt{\rho_0}$ such that $\rho_0={\rm Tr}_B\wt{\rho_0}$, and then the entropy is given by the entanglement $S(\rho_0)=E(\wt{\rho_0}|A:B))$.
Finally, proposition \ref{entG} gives us:
$$
E(\wt{\rho_0})=\sum_{j=0}^n \sum_{a_j,b_j} w^{(j)}_{a_j, b_j} \log(2j+1),
\qquad\textrm{with}\quad
\sum_{j=0}^n \sum_{a_j,b_j} w^{(j)}_{a_j, b_j}=1.
$$
Obviously, the maximal entanglement will  be when $w^{(n)}=1$ and
$w^{(j)}=0$ for all $j<n$: Soraya must use the highest possible
spin $j=n$ to encode her messages. Then the quantum capacity is
given by the entanglement, so that ${\rm QCap}=\log(2n+1)$, as
obtained in \cite{crypt}.

Finally, \cite{crypt} notices that the classical capacity of this secured quantum channel is three times its quantum capacity. This factor 3 is exactly the one between the classical correlations and quantum correlations which we obtained earlier in equation \Ref{qmi3}.

To conclude this section, we have shown that this particular protocol of quantum communication without shared reference frame allows to translate directly physical properties of the quantum black hole, like its entropy or entanglement, to quantum information notions such as classical and quantum channel capacities. We hope that such a dictionary  could be further developed.

\section{The spin-1 black hole model and generalization}

We now describe higher spin models, where we assume that the
horizon is made of $n$ elementary patches of a fixed spin $s$. We
start by analyzing the 1-spin case and then we generalize the
result about the entropy of the general case. These models can be
interpreted  as coarse-grained models of the fundamental
$1/2$-spin model when the observer counting the degrees of freedom
does not have access to a resolution finer than the unit area $a_s$
set by the elementary patch of spin $s$.

In the spin-1 case, we are interested into the decomposition of
multiple tensor products of $V^1$:
\be
\bigotimes^{n}V^1
\,=\,
\bigoplus_{j=0}^n \d1_j^{(n)}\,V^j.
\ee
We show below how to compute the degeneracies $\d1_j^{(n)}$ in terms of random walk.
Then we explain how the random walk calculation extends to the general case of the decomposition of multiple tensor products of $V^s$:
$$
\bigotimes^{n}V^s
\,=\,
\bigoplus_{j=0}^n \dk_j^{(n)}\,V^j.
$$
This allows us to prove that the black hole entropy
$S^{(s)}(n)\equiv\log(\dk_0^{(n)})$ always
 have a first term proportional to the horizon area and then a logarithmic correction with the same $-3/2$
 factor. Similarly, Eq.~(\ref{qmi3}) that compares quantum mutual
 information $I_\rho(A:B)$ for an arbitrary bipartite splitting of
 the horizon spin network with the entanglement between its halves
 holds also in this case.

This claim was already made in \cite{spinone}. However, there, the authors start with the number
of conformal blocks for the q-deformed $\SU(2)$. We believe that our approach is simpler and more
transparent, and that the random walk analogy allows a straightforward generalization to a large
class of gauge groups (other than $\SU(2)$). For example, the same results with the universal $-3/2$
 factor directly applies to the supersymmetric extension $\osp(1|2)$, which is the relevant group for
 the loop quantization of $N=1$ 4d supergravity.

\subsection{Random walk analogy, entropy renormalisation and universal log correction}

Following the same logic as for the $1/2$-spin model, we describe the decomposition of the tensor product
$\otimes^{n}V^s$ iteratively. Since for an arbitrary spin $j$ representation we have
$$
V^s\otimes V^{j_0} =\bigoplus_{j=|j_0-s|}^{j_0+s}V^j,
$$
the iterative process can be thought as a random walk. From the spot $j_0$, one can go in one step with equal probability to any spot from $j_0-s$ to $j_0+s$, as long as $j_0$ is larger than the fixed spin $s$. When $j_0$ is smaller than $s$, then we run into the wall at the origin $j=0$. We formalize this in the following statement:
\begin{prop}
\label{RW}
The degeneracy of the irreducible spin representation $V_j$ in the decomposition of the tensor product $\otimes^{n}V^s$ is the number of returns to the spot $j$ after $n$ iterations of a random walk allowing jumps of length up to $s$ with a wall/mirror at $j=0$. This truncated random walk is such that the number of returns can be simply computed from a standard random walk allowing jumps of length up to $s$ with no obstacle at $j=0$:
\be
\dk_j^{(n)}=\rwm_n^{(s)}(j)=\rw_n^{(s)}(j)-\rw_n^{(s)}(j+1).
\ee
Then the number of returns of the standard random walk can be computed as an integral:
\be
\rw_n^{(s)}(j)
=\f{1}{2\pi}\int_{-\pi}^{+\pi} d\theta\,e^{-ij\theta}
\left(e^{-is\theta}+..+1+..+e^{+is\theta}\right)^n,
\label{randomint}
\ee
for $s$ is an integer and
\be
\rw_n^{(s)}(j)
=\f{1}{2\pi}\int_{-\pi}^{+\pi}d\theta\, e^{-i(2j)\theta}
\left(e^{-i(2s)\theta}+e^{-i(2k-2)\theta}+..+e^{-i\theta}+e^{i\theta}+..+e^{+i(2s)\theta}\right)^n,
\ee
when $s$ is a half-integer.
\end{prop}

As in the $1/2$-spin case, it is actually possible to derive these integral formulas directly using the characters of $\SU(2)$:
\be
\dk_j^{(n)}\,=\,\int dg\, \chi_j(g)\left(\chi_s(g)\right)^n
\,=\,\f 2\pi\int_0^\pi d\theta\,\sin\theta\,\sin(2j+1)\theta\,
\left(\f{\sin(2s+1)\theta}{\sin\theta}\right)^n.
\ee

For $s=1$, we get explicit formulas of $\rw_n^1(0)$  and
$\rwm_n^1(0)$ as sums:
\be
\rw_n^1(0)=\sum_{l=0}^{\left\lfloor\f n2\right\rfloor}\f{n!}{l!l!(n-2l)!},
\ee
\be
\rwm_n^1(0)=\sum_{l=0}^{\left\lfloor\f n2\right\rfloor}\f{n!}{l!l!(n-2l)!}
-\sum_{l=0}^{\left\lfloor\f
{n-1}2\right\rfloor}\f{n!}{l!(l+1)!(n-2l-1)!},
\ee
where $\lfloor y\rfloor$ is the integer part of $y$.
Distinguishing the cases with $n$ is odd or even, one can further
simply the expression of the truncated random walk\footnotemark.
\footnotetext{For example when $n=2m+1$, we have:
$$
\rwm_n^1(0)=\sum_{l=0}^{m}\f{n!}{l!l!(n-2l)!}\f{3l-n+1}{l+1}.
$$}
We can directly evaluate the asymptotics of these numbers of return using Stirling formula. Indeed for large $n$, letting $x\equiv 2l/n\in[0,1]$, the sums can be approximated by the integrals:
$$
\rw_n^1(0)\sim \f{1}{2\pi}\int_0^1 dx\,\f{1}{x\sqrt{1-x}}e^{n\phi(x)},
\qquad
\rwm_n^1(0)\sim\f{1}{2\pi}\int_0^1 dx\,\f{3x-2}{x^2\sqrt{1-x}}e^{n\phi(x)},
$$
where the exponent is given by
$$
\phi(x)=x\ln 2 -x\ln x -(1-x)\ln (1-x).
$$
It is straightforward to check that $\phi$ admits a unique maximum at $x=2/3$ for which $\phi(2/3)=\ln 3$ and
$\phi''(2/3)=-9/2<0$. We then directly extract the asymptotical behavior of the integrals using the saddle point approximation. Since $1/x\sqrt{1-x}$ doesn't vanish at $x=2/3$ we get:
\be
\rw_n^1(0)\propto\f{3^n}{\sqrt{n}},
\ee
while since $(3x-2)/x^2\sqrt{1-x}$ vanishes at $x=2/3$ but has a non-vanishing second derivative\footnotemark, we get:
\be
\rwm_n^1(0)\propto\f{3^n}{n\sqrt{n}}.
\ee
\footnotetext{Let us consider the integral $\int_0^1 dx\, f(x)e^{n\phi(x)}$. Assuming that $\phi$ has a unique maximum at $x_0\in]0,1[$, we write $\phi(x)=a-b(x-x_0)^2$ with $b>0$. Then we expand $f$ around $x_0$ and doing the change of variable to $y=(x-x_0)\sqrt{n}$, which gives the asymptotic behavior of the integral as:
$$
\f{e^{na}}{\sqrt{n}}\int_{-\infty}^{+\infty}dy\,
\left(f(x_0)+f'(x_0)\f{y}{\sqrt{n}}+\f12f''(x_0)\f{y^2}{n}+..\right)e^{-by^2}.
$$
It is clear that when $f(x_0)$ vanishes, the first derivative $f'(x_0)$ doesn't play
 into role in the approximation since $\int dy\, ye^{-by^2}=0$, so that the first relevant term is the second derivative $f''(x_0)$.}

A faster and easier calculation is using the saddle point approximation directly
on the integral expression given in the previous proposition \ref{RW}. Indeed:
\be
\rw_n^1(0)=\f{1}{2\pi}\int_{-\pi}^{+\pi}d\theta\,(1+2\cos\theta)^n
=\f{1}{2\pi}\int_{-\f{2\pi}{3}}^{+\f{2\pi}{3}}d\theta\,e^{n\ln(2\cos\theta+1)}
+\f{(-1)^n}{2\pi}\int_{-\f{\pi}{3}}^{+\f{\pi}{3}}d\theta\,e^{n\ln(2\cos\theta-1)}.
\ee
Both exponents $(2\cos\theta+1)$ and $(2\cos\theta-1)$ have a unique maximum at $\theta=0$.
 Thus the first term will contribute $3^n/\sqrt{n}$ to the asymptotic behavior while the second term will only give a term of the order of $1/\sqrt{n}$. In the end, we simply recover $\rw_n(0)\sim 3^n/\sqrt{n}$.
Now,
\be
\rwm_n^1(0)=\f{1}{2\pi}\int_{-\pi}^{+\pi}d\theta\,(1-e^{-i\theta})(1+2\cos\theta)^n
=\f{1}{2\pi}\int_{-\pi}^{+\pi}d\theta\,(1-\cos\theta)(1+2\cos\theta)^n.
\ee
As $(1-\cos\theta)\underset{\theta\approx0}\sim\theta^2/2$, the saddle point approximation directly yields  $\rwm_n(0)\sim 3^n/n\sqrt{n}$.

This proof can be straightforwardly adapted to the generic
spin-$s$ case, so that we get:
\begin{prop}
We have the following asymptotical behavior of the degeneracy of the trivial representation
 $V^0$ in the decomposition of the tensor product $\otimes^{n}V^s$:
\be
\rw_n^{(s)}(0)\underset{n\arr\infty}{\propto} \f{(2s+1)^n}{\sqrt{n}},
\qquad
\dk^{(n)}_0=\rwm_n^{(s)}(0)\underset{n\arr\infty}{\propto} \f{(2s+1)^n}{n\sqrt{n}}.
\ee
\end{prop}
Since the degeneracy of the trivial representation is simply the dimension of the intertwiner space
 i.e. of the Hilbert space of horizon states, we derive the entropy law:
\be
S^{(s)}_n\sim n\ln(2k+1)-\f{3}{2}\ln n +..,
\ee
where we recognize the first term proportional to the horizon area and the now universal $-3/2$ factor in front of the logarithmic correction.

Here, we showed that the $3/2$ correction factor comes directly from the analogy with the random walk and does not depend on the particular model (choice of spin for the black hole elementary surfaces) that we  chose. This proves that the $3/2$ factor is actually invariant under coarse-graining i.e under the choice of area unit that we use to probe the horizon of the black hole: this is therefore a universal factor and rigid prediction from these particular models of quantum black holes.

Moreover, we can refine our evaluation of the degeneracies $\dk_j^{(n)}$ pushing further the approximations of the integrals giving $\rwm$:
$$
\dk_j^{(n)}\underset{j\ll n}{\sim}
\f{3}{8}\sqrt{\f{3}{2\pi}}\f{(1+2s)^n}{[s(1 + s)]^{5/2}}
\f{4s(1+s)n-3 -6j(1+j)}{n^{5/2}}(1+j)
\underset{n\arr\infty}{\sim}\f{(1+2s)^n(1+j)}{(s(1+s)n)^{3/2}}.
$$
This allows us to show the following statement.
\begin{prop}
 We have the following asymptotical behavior of the maximal
dimensionality  $\dk^{(n)}_{j_{\max}}={\max}_j\ \dk^{(n)}_{j} $
 of the representation spaces $V^{j}$ in the decomposition of the tensor
product $\otimes^{n}V^s$:
\be
j_{\max}\underset{n\arr\infty}{\propto}\sqrt{n},\qquad
\dk^{(n)}_{j_{\max}}\underset{n\arr\infty}{\propto}
\f{(2s+1)^n}{n}.
\ee
The proportionality coefficients depend on $s$.
\end{prop}
The calculations leading to these results can be found in Appendix B.


\subsection{Elementary scale and irreducibility} \label{elscale}

The structure of spacetimes as given by LQG is fundamentally
discrete.  Hence the diffeomorphism symmetry that is defined on a
smooth manifold is replaced by a discrete symmetry on a spin
network
\cite{BHlee}. The Hilbert space of LQG is spanned by the spin networks, and it
carries a representation of the permutation group. Any
 representation of the permutation group of $n$ objects $S_n$ can be either
faithful or irreducible. Moreover, for large $n$ the squared
dimension of  $(\C^k)^{\otimes n}$ is smaller than $n!$, so the
faithful representation of the permutation group on the spin
network  any size is impossible. There is no a priori reason
\cite{BHlee} to require that all the edges of a spin network are labeled by the
same spin $s$, since it is not necessary for building
representations of the permutation group. However, we can see that
the requirement of the irreducibility of the representation is
equivalent to describing  spin networks only in terms of the
fundamental representation $s=\half$. Namely
\cite{good,che}, the Shur's duality that allows to decompose
$(\C^N)^n$ as a direct sum of tensor products of the irreps of
SU($N$) and $S_k$ for the direct product of U($N$) holds only when
the dimension of the elementary space is $N$. For SU(2) it takes
the form
\be
\bigotimes^{2n}\C^2\cong\bigoplus_{j=0}^{n} V^j\otimes
\sigma_{n,j},
\ee
where $V^j$ is the irreducible spin-$j$ representation  of
$\SU(2)$, and the irreps $\sigma_{n,j}$  correspond to the
partition $[n+j,n-j]$ of $2n$ objects
\cite{wkt}. There is no similar structure for the direct product
of different vector spaces on the left hand side, and the
isomorphism in the case of the direct product of $\C^N$ involves
the irreducible representations of SU($N$).

\section{On area renormalisation in LQG}
\label{renormalisation}

Let us look at a generic surface, made of $2n$ elementary patches
of spin-$1/2$. Unlike  the previous case we allow for open
surfaces. Then the Hilbert space of surface states is simply the
tensor product
 $\otimes^{2n}V^{1/2}$. The precise state describes how the geometry of the surface, the way
 it is folded. It is the surrounding spin network, in which the surface is embedded, which provides
 the information on the surface state and the way that the surface is precisely embedded in the 3d space.

Now the surface can be folded any way at the microscopic scale. Then depending on how the surface is folded,
 an observer at a larger will see a smooth surface which is only a approximation of the ``real" surface
 and will measure an area smaller than the full microscopic area $\aam=2n a_{1/2}$.

\medskip

Let us start by the case when we completely coarse-grain the surface state.
That is we decompose $\otimes^{2n}V^{1/2}=\oplus_j^n d_j^{(n)} V^j$,
and the total spin $j$ will describe the macroscopic area of the completely coarse-grained surface. For example, for $2n=2$, $V^{1/2}\otimes V^{1/2}=V_0\oplus V^1$ and we have two coarse-grained area states: $j=0$ which corresponds to the surface folded on itself and $j=1$ which corresponds to the open surface with the two elementary patches side by side.

For large $n$, one can describe the completely
 coarse-grained state for the trivial state of geometry i.e. when
  we do have no information at all the state of the surface.
  Then the most probable macroscopic area is given by the spin $j$ with
  the largest degeneracy $d_j^{(n)}$. Given the explicit expression of
  these degeneracy factors as $d_j=C_{2n}^{n+j}-C_{2n}^{n+j+1}$, we easily write that
$$
\f{d_{j+1}^{(n)}}{d_j^{(n)}}=\f{(n-j)(2j+3)}{(n+j+2)(2j+1)},
$$
and identify the maximal entropy for:
\be
j_{\max}^{\left(\f{1}{2}\right)}=\sqrt{\f{n}{2}}.
\ee
One can reproduce the same result by using the Stirling formula approximation\footnotemark {} to $d_j$ for large $n$.
\footnotetext{Letting $x=j/n\in[0,1]$, we have
\be
d_j^{(n)}\approx
d^{(n)}(x)=\f{2^{2n}}{\sqrt{n\pi}}f(x)e^{-n\phi(x)},
\label{djasym}
\ee
with
$$
f(x)=\f{1}{\sqrt{1-x^2}}\f{2x+\f{1}{n}}{1+x+\f{1}{n}}, \qquad
\phi(x)=(1-x)\ln(1-x)+(1+x)\ln(1+x).
$$
The maximum is given by the equation $n\phi'(x)=(\ln f)'(x)$. For
$x_{\max}\approx 0$, we get $x_{\max}\sim 1/\sqrt{2n}$. Moreover
if we prefer to look at the maximum of $(2j+1)d_j^{(n)}$, then one
simply multiply the function $f$ by $2x+1/n$ and then the maximum
is located at $x_{\max}\sim 1/\sqrt{n}$.} This square-root law is
natural from the point of view of the random walk analogy and it
is actually true in any $s$-spin model, when looking at the
maximal degeneracy in the decomposition of the tensor product
$\otimes^{2n}V^s$ (see Appendix B for details).

However, as soon as we have a non-trivial geometry state and have more information of the tensor
 product state corresponding to the surface at the microscopic level, we do not expect the macroscopic
 area to follow this rule anymore. Nevertheless, we would like to insist on the fact that the maximal area
 given by the microscopic area $2n a_{1/2}$ is totally improbable at the macroscopic level: its degeneracy
  is simply 1 while typical degeneracies grow as $2^{2n}/n\sqrt{n}$ and the maximal degeneracy
  as $2^{2n}/n$. Therefore we believe that the study of area reorganization in LQG will not be
   as straightforward as assuming a simple linear rescaling of the microscopic area.

\medskip

More generally, we do not need to go directly at the completely coarse-grained level.
Physically, we are interested by the whole coarse-graining process: for example going
 from the fundamental scale set the $1/2$-spin to a larger semi-classical scale given
 a fixed large $s$-spin. A simple calculation one can do considering $2n$ qubits is
 to take a subset of $m$ qubits and compute the reduced density matrix of the subset
  from the state (possibly mixed) of the full surface. Then one will know the most
   likely area of this subsurface formed by the considered $2k$ qubits. Pushing this
   set-up further, we can partition the full surface into $p$ bigger patches formed
   by $n/p$ $1/2$-spins and compute the reduced density matrices of each bigger patch
   and thus obtain a coarse-grained description of the surface. A first step would be
   to do this calculation in the framework of the totally mixed state corresponding to
   the black hole state.

Indeed, a reduced density matrix for
 a  $2k$ qubit patch is
 \be
\rho_{(2k)}=\frac{1}{N}\sum_{j=0}^{k}\sum_{a_j=1}^{d_j^A}d_j^B\rho_{j,a_j}.
 \ee
At this stage of the analysis it is not clear which prescription
to calculate a coarse-grained value of $j$  should be used. There
are three reasonable options that all lead to $j_{\rm{coarse}}\sim
\sqrt{k}$, while the numerical factors are of the order of unity.
\begin{itemize}
\item The expectation value of the patch's spin is
 \be
 \6 J\9=\tr J\rho_{(2k)},
 \ee
where $J=\sqrt{{\bf J}^2}$ or any other suitable function of ${\bf
J}$. Since all the density matrices are diagonal in the angular
momentum basis, this expectation value can be calculated as
\be
\6 J\9=\sum_j^k J(j) p_j^{(n,k)}, \qquad
p_j^{(n,k)}=d^{(2k)}_jd^{(2n-2k)}_j/N,\label{Jsum}
\ee
and, e. g.,  $J=\sqrt{j(j+1)}$. We calculate the expectation value
of $J$ by replacing the sum in Eq.~(\ref{Jsum}) by the integral.
Using the asymptotic expansion of Eq.~(\ref{djasym}) the
normalization of the probability distribution $p_j^{(n,k)}$ is
fixed by
\be
\int_0^\infty x^2e^{-(\mu+1) k x^2}=N',
\ee
where $\mu\equiv k/(n-k)\approx k/n$, all the constants where
absorbed in $N'$ and $x_{\max}\approx 1/\sqrt{2k}\rightarrow 0$
was used. Taking $J\approx j$ the expectation value becomes
\be
\6J\9\approx\6 j\9=\frac{k}{N'}\int_0^\infty x^3e^{-(\mu+1) k
x^2}=\sqrt{k}\frac{2}{\sqrt{\pi}}.
\ee
Since for large spins the elementary area is approximately
proportional to the spin,
\be
\f{\aaM}{\aam}\approx \f{2}{a_{1/2}\sqrt{\pi}}\sqrt{\f{p}{n}},
\ee
as illustrated on Fig.~2 below.
\begin{figure}[htbp]
\epsfxsize=0.45\textwidth
\centerline{\epsffile{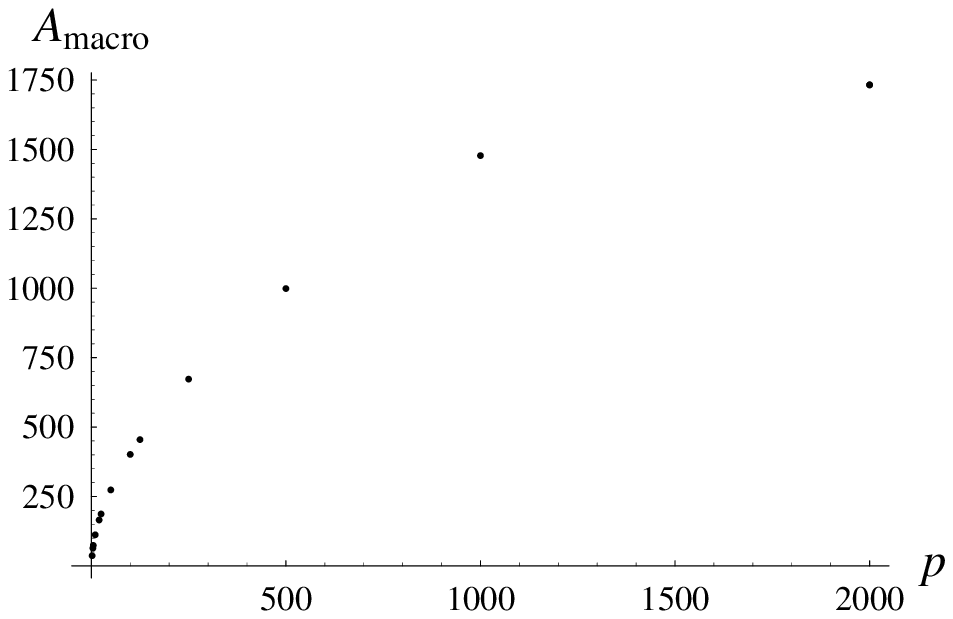}{\hspace{-3cm}\raisebox{1cm}{\epsfxsize=0.25\textwidth\epsffile{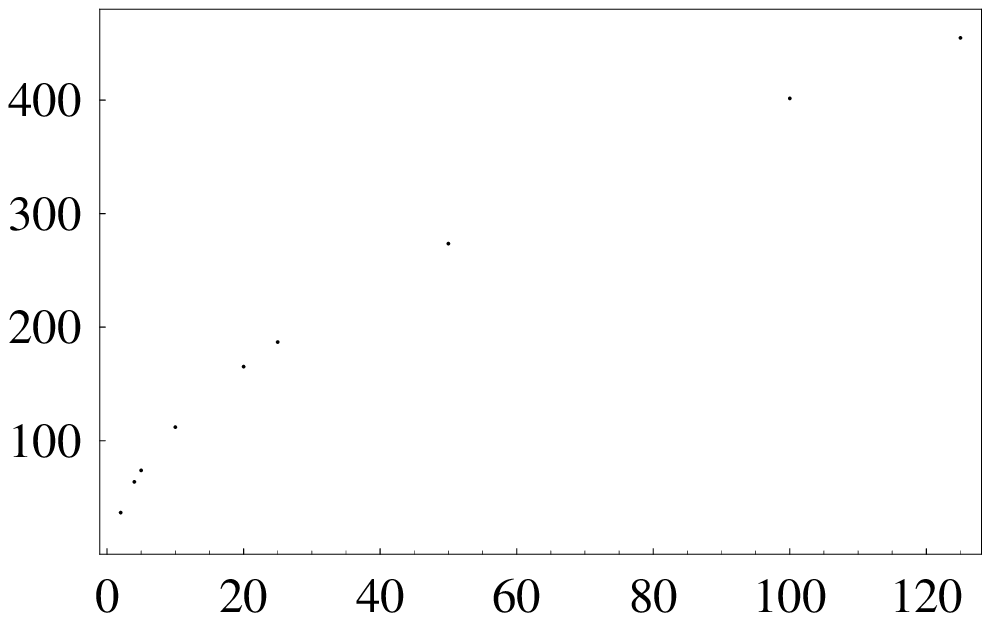}}}}
\caption{\small{The macroscopic area of the $2n=2000$ qubit horizon as a function of the number of patches, $p=2,4,5,\ldots,2000$.
The square-root area spectrum is assumed, $a_j=\sqrt{j(j+1)}$, and
$\6J\9$ is taken to represent the patch.}}
\end{figure}

\item  Considering $\6 J\9$  as an
expectation values of a spin measurement,
 the most likely spin value $j$ that occurs in such a measurement is
$j_{\max}^{(n,k)}$ that maximizes the probability distribution
$p_j^{(n,k)}$.  When $1\ll k$ the spin of the highest degeneracy
 representation is given by Eq.~(\ref{jmaxnm}), and for $k\ll n$
 it reduces to
\be
 j_{\max}=\sqrt{k}.
\ee

\item The most likely state $\ket{j,j_z}$ (again summing over
all the degeneracy labels) corresponds to $\bar{\jmath}$ that
maximizes
\be
p_{j,j_z}^{(n,k)}=\frac{p_j^{(n,k)}}{2j+1}.
\ee
 The value of $\bar{\jmath}$ can be determined similarly to $j_{\max}$, and for $1\ll
 k\ll n$ it is
 \be
 \bar{\jmath}=\sqrt{k/2}.
 \ee
\end{itemize}

\noindent For all the three choices of the coarse-grained $j$ the area of
the patch is
\be
\aa_{(2k)}=a_{1/2}\alpha\sqrt{k},
\ee
where $\alpha\sim1$ depends on the coarse-graining procedure. As a
result, the coarse-grained area of the surface is
\be
\aaM=p\aa_{(2k)}=a_{1/2}\alpha\sqrt{n p}.
\ee

In the end, it would be interesting to generalize these area
renormalisation calculations to more generic situations in loop
quantum gravity, and thereby underline the distinction between the
microscopic geometry and the semi-classical measurable and
observable geometry.

\section{Conclusions and Outlook}

We have studied simple models for black holes in the framework of
loop quantum gravity. We consider the black hole horizon as a
quantum surface made of a certain number $n$ of identical
elementary surfaces. These elementary quantum surfaces are
mathematically modeled as spin-$s$ systems, while the whole
horizon must be a singlet state (or intertwiner)
 in the tensor product of these $n$ spins $s$. Actually our framework applies to generic quantum closed surfaces as soon
 as we coarse-grain the interior geometry.

We have computed the entropy of black holes in this context and recovered the usual entropy with a logarithmic correction. The factor in front of that correction is actually $-3/2$ whatever the spin $s$ which we choose. We also analyzed the entanglement between parts of the horizon. More technically, we computed the entanglement for splitting of the horizon into two parts of arbitrary sizes. We have shown that this entanglement is responsible for the logarithmic correction to the entropy law. More precisely, the logarithmic correction is  the total correlation between the two parts of the horizon and is equal to three times the entanglement. Moreover, we also point out a potential relation with the evaporation process, which allows us to recover at first order the Hawking evaporation formula. From our perspective, quantum black holes appear at the interplay between quantum gravity and quantum information, and we show how all the objects appearing in our analysis of the black hole can be translated to the framework of quantum communication without shared reference frame.

Finally, we have used all the developed technology to illustrate a concept of renormalisation
of areas in loop quantum gravity: given a surface at the microscopic level, it might be folded
 in ways not observable to a macroscopic observer, so that the macroscopically observed area
  is different to the microscopic area postulated by loop quantum gravity.
   And we formulate a square-root law for the renormalisation flow of the area.
    We see a few possible development of the present work:
\begin{itemize}
\item One can try to introduce the dynamics into these simple black hole models,
using the quantization of the ADM energy in loop quantum gravity. This would allow
to study the evolution of the entanglement, potentially look at the entanglement decay and the evaporation process.
\item One can try to take into account the rotation of black holes in our models.
 However, this means to introduce more objects, relative to which we would define the
 rotation of the black hole. This would need to model the exterior of the black holes and think about spin network geometries whose semi-classical regimes are approximately the Schwarzschild metric.
\item It would be interesting to push further our analysis of area renormalisation, and study this concept in more complicated situation that the black hole state which turned out to be rather simple. We could also look at the renormalisation of other geometrical quantities such as the volume.
\item Finally, we can push further the use of entanglement and correlations as tools to probe
 the geometry of spin networks in LQG. One can naturally speculate on a link between the notion
  of distance and the actual correlations between subsystems of a given spin networks, similarly
  to what happens in spin lattices. Work on this topic will be reported elsewhere \cite{QIQG2}.
\end{itemize}

\section*{Acknowledgments}
We would like to thanks Laurent Freidel, Lee Smolin and Florian Girelli for their interest into
 the present work.
We are also grateful to Steve Bartlett for a very very long discussion on reference-frame-based
cryptography protocols in quantum communication during some warm days in Sydney.


\appendix

\section{Entanglement of the black hole state}

Here we prove the Proposition \ref{BHentprop} giving the exact
formula for the entanglement of the black hole state. In the basis
we adopted the states $|j,a_j,b_j\9$ are given be the
bi-orthogonal (Schmidt) decompositions
\cite{per}, so their reduced density matrices are diagonal.
Moreover, the only nonzero part of $\rho_{j,a_j}$ is a sequence of
$2j+1$ terms $1/(2j+1)$. Using the decomposition of $A$ into $V^j$
and $D^j$ subspaces,
\be
\rho_{j,a_j}=\frac{1}{2j+1}\1_{V^j}\otimes|a_j\9\6a_j|_{D^j}
\ee
 For the future convenience we introduce $\rho_j=\1_{V^j}/(2j+1)$. Two   matrices $\rho_{j,a_j}$
 and $\rho_{l,a_l}$ have orthogonal  supports if one or both of their indices
are different. The reduced density matrix $\rho_A=\tr_B
\rho$ is
\be
\rho_A=\frac{1}{N}\sum_j \sum_{a_j}d_j^B \rho_j\otimes|a_j\9\6a_j|_{D^j}.
\ee
Since the von Neumann entropy of the reduced density matrices
 $\rho_{j}$ satisfies
\be
S(\rho_{j, a_j})=S(\rho_{j})=\log(2j+1), \label{ent0}
\ee
the weighted average of the entanglement in this decomposition is
\be
S_E(\rho)=\frac{1}{N}\sum_{j=0}^{m}\sum_{a_j=1}^{c_j^A}\sum_{b_j=1}^{c_j^B}S(\rho_{j,a_j})
=\frac{1}{N}\sum_{j=0}^{m}d_j^Ad_j^B\log(2j+1) \label{ent1}.
\ee
All pure states $|\Psi_\alpha\9$ that appear in alternative
decompositions of $\rho$ 
ought to be some linear combinations of the states $|j,a_j,b_j\9$,
\be
|\Psi_\alpha\9=\sum_{j, a_j, b_j}
c_{\alpha,ja_jb_j}|j\9_{AB}\otimes|a_j\9_{D^j_A}\otimes|b_j\9_{D^j_B}.
\ee

The diagonal form of $\rho$  forces the coefficients
$c_{\alpha,ja_jb_j}$ to satisfy the normalization condition
\be
\sum_\alpha
w_\alpha
c_{\alpha,ja_jb_j}c^*_{\alpha,la_lb_l}=\frac{1}{N}\delta_{jl}\delta_{a_l
a_j}\delta_{b_j b_l}. \label{wnorm}
\ee
The reduced density matrices
$\rho_A(\alpha)=\tr_B|\Psi_\alpha\9\6\Psi_\alpha|$  are
\be
\rho_A(\alpha)=\sum_j\sum_{b_j}\sum_{a_j,a'_j}c_{\alpha,ja_jb_j}c^*_{\alpha,ja'_jb_j}
\rho_j\otimes|a_j\9\6a'_j|_{D^j_A}.
\ee
Introducing
\be
\lambda_{\alpha,ja_ja'_j}=\sum_{b_j}c_{\alpha,ja_jb_j}c^*_{\alpha,ja'_jb_j},
\quad \pi_j(\alpha)=\sum_{a_j} \lambda_{\alpha, j, a_j a_j},
\ee
we rewrite the reduced density matrix of
$|\Psi_\alpha\9\6\Psi_\alpha|$ as
\be
\rho_A(\alpha)=\sum_j \pi_j(\alpha) \rho_j\otimes \Lambda_j(\alpha),
\ee
where the fictitious density matrix $\Lambda$ was introduced on
the degeneracy subspace $D^j_A$,
\be
\Lambda_j(\alpha)=\frac{1}{\pi_j(\alpha)}\sum_{a_ja'_j}\lambda_{\alpha,ja_ja'_j}|a_j\9\6a'_j|_{D^j_A}.
\label{lambda}
\ee
From the orthogonality  of the matrices $\rho_{j, a_j}$ and
Eq.~(\ref{wnorm}) it is easy to see that
\be
\sum_\alpha w_\alpha\lambda_{\alpha,ja_ja'_j}=\frac{1}{N}d_j^B\delta_{a_j
a'_j}.
\ee

The weighted average of the entanglement of the decomposition
$\{\Phi_\alpha\}$ is $
\6 S(\{\Phi_\alpha\})\9\equiv\sum_\alpha w_\alpha S(\rho_A(\alpha))$.
From Eq.~(\ref{lambda}) and the concavity property of entropy
\cite{per,wehrl} it follows that
\begin{widetext}
\be
\6 S(\{\Phi_\alpha\})\9\geq \sum_{\alpha, j}w_\alpha
\pi_j(\alpha)\left[S(\rho_j)+S(\Lambda_j(\alpha))\right]\geq\sum_{\alpha, j}w_\alpha
\pi_j(\alpha) S(\rho_j)=\frac{1}{N}\sum_j d^A_j d^B_j S(\rho_j).
\label{qed} \ee
\end{widetext}
Hence $S_E(\rho)$ is indeed the entanglement of formation. The
equality of all measures of entanglement for the black hole states
is proven in \cite{lit1}.

\section{Evaluation of the degeneracies $\dk_j^{(n)}$ for arbitrary spin}

We consider the integer spin. The half-integer spin is treated in
exactly the same way. From Eq.~(\ref{randomint}) it follows that
\be
\rwm_n^s(j)=\f{1}{2\pi}\int_{-\pi}^{\pi}\!d\theta
f(j,\theta)e^{n\log F(s,\theta)},
\ee
where $f(j,\theta)=\cos(j\theta)-\cos[(j+1)\theta]$ and
\be
F(s,\theta)=
\f{2\cos\left(\f{s\theta}{2}\right)\sin\left(\f{(1+s)\theta}{2}\right)}
{\sin\left(\f{\theta}{2}\right)}-1.
\ee
For a fixed $j$ and $n\rightarrow\infty$ the saddle point method
gives the following estimate of the degeneracy:
\be
\dk_j^{(n)}\approx\f{3}{8}\sqrt{\f{3}{2\pi}}\f{(1+2s)^n}{[s(1 + s)]^{5/2}}
\f{4s(1+s)n-3 -6j(1+j)}{n^{5/2}}(1+j). \label{dkjf}
\ee
Hence
\be
\dk_j^{(n)}\underset{n\arr\infty}{\propto}\f{(1+2s)^n(1+j)}{(s(1+s)n)^{3/2}}
\label{dsjb}
\ee
Having in mind the search for $j_{\max}$,
$\dk_{j_{\max}}^{(n)}=\max$, in deriving (\ref{dkjf}) we expanded
$f(j,\theta)$ as $f(j,\theta)\approx a\theta^2+b\theta^4$. Indeed,
solving $d\dk_j^{(n)}/dj=0$ we get
\be
j_{\max}\approx \f{\sqrt{8k(1+s)n-3}}{6}-\f{1}{2}, \qquad
\dk_{j_{\max}}^{(n)}\underset{n\arr\infty}{\propto}\f{(1+2k)^n}{n}.\label{dsmax}
\ee
A comparison with the exact results for the degeneracies for
$s=\half$ and $s=1$ shows an excellent agrement with the
asymptotics for $j\lesssim\sqrt{n}$. In both cases numerics show that the asymptotic
estimate of $j_{\max}$ given above differs from its exact value by a factor of
$\sqrt{2/3}$ and that the maximal degeneracies are of the same order of
magnitude.
\section{Entanglement calculations}

 Since Eq.~(\ref{ent1}) gives the entanglement of formation of
 $\rho$,
 it is necessary to find the
multiplicities of the spin-$j$ subspaces. For $s=\half$ the result
is
\be
d^{(2k)}_j={2k\choose k+j}\frac{2j+1}{k+j+1},
\ee
where instead of the subspace labels $A,B$ we write the number of
qubits. The multiplicities satisfy the following normalization
conditions:
\be
\sum_{j=0}^{k}d_j^{(2k)} d_j^{(2n-2k)}=N, \qquad \sum_{j=0}^k
d_j^{(2k)}={2k\choose k}, \qquad \sum_{j=0}^k
d_j^{(2k)}(2j+1)=2^{2k}. \label{djform}
\ee

We start the entanglement evaluation  by considering  two equal
subsystems of $n$ qubits each. In this case $d_j^{A,B}\equiv d_j$,
so
\be
E(\rho|n:n)=\frac{1}{N}\sum_{j=0}^{n/2}d_j^2\log(2j+1).
\ee
In the leading order
\be
E(\rho|n:n)\sim\log(2j_{\max}^{(n/2)}+1)\cdot 1, \label{approA}
\ee
where the coefficient $d_{j}^{(2k)}$ reaches its maximal value at
\be
j=j_{\max}^{(k)}\approx\half\sqrt{2k+2}-1,\label{jmax1}
\ee
so
\be
E(\rho|n:n)\sim\half\log n. \label{entc}
\ee
Numerical simulations  show that $E(\rho|n:n)-\half\log n\approx
0.0183$

In a generic case of $2k:2n-2k$ splitting for $1\ll k\leq n/2$ the
sum in Eq.~(\ref{ent1}) can be evaluated similarly to
Eq.~(\ref{approA}).  In this case
\be
j_{\max}^{(n,k)}\approx\frac{1}{2}\sqrt{\frac{2+3n+4k(n-k)}{n+1}}-1,
\label{jmaxnm}
\ee
which reduces to the result of Eq.~(\ref{jmax1}) for the $n:n$
splitting. It is also interesting to note the fraction of the
unentangled states in the decomposition of $\rho$
\be
s^{(2k)}_0\equiv\f{d^{(2k)}d^{(2n-2k)}}{N},
\ee
goes down with $k$, $1\ll k\leq n/2$,
\be
 s^{(2k)}_0\sim
\frac{{1}}{\sqrt{\pi}}\left(\frac{n}{k(n-k)}\right)^{3/2}.
\ee

The quantum mutual information was introduced in Sec.~2.2.  The
reduced density matrix of a smaller block for $2k:2n-2k$ splitting
is
\be
\rho_{(2k)}=\frac{1}{N}\sum_{j=0}^{k}\sum_{a_j=1}^{d_j^A}d_j^B\rho_{j,a_j},
\ee
and it entropy is
\be
S(\rho_{(2k)})=-\sum_{j=0}^{k}(2j+1)d_j^A
\left(\frac{d_j^B}{N}\frac{1}{2j+1}\right)\log\left(\frac{d_j^B}{N}\frac{1}{2j+1}\right).
\ee
In particular, for the  $n:n$ decomposition
\be
S(\rho_{(n)})=E(\rho)+\sum_{j=0}^{n/2}\frac{d_j^2}{N}\log{\frac{N}{d_j}}
=E(\rho)+\log{N}-\sum_{j=0}^{n/2}\frac{d_j^2}{N}\log{d_j}.
\ee
Again we approximate
\be
\sum_{j=0}^{n/2}\frac{d_j^2}{N}\log{d_j}\sim \sum_{j=0}^{n/2}\frac{d_j^2}{N}\log{d_{j_{\max}}}
=\log{d_{j_{\max}}},\label{entapc}
\ee
where $\log d_{j_{\max}}\sim n\log2-\log n$, so
\be
S(\rho_{(n)})\sim E(\rho|n:n)+n\log2-\half\log{n}+\ldots\sim
n\log2.
\ee
As a result,
\be
I_\rho(n:n)\equiv 2 S(\rho_{(n)})-S(\rho)\sim 3E(\rho|n:n).
\label{cmi}
\ee

Using the results of Appendix B it is possible to show that the
above relation is true for any fundamental spin $s$. Indeed, from
Eq.~(\ref{dsjb}) it follows that $N^{(s)}\equiv \dim\hh$ satisfies
\be
\log N^{(s)}=\log{\dk_j^{((n)}(0)}\sim n\log(1+2k)-\f{3}{2}\log n,
\ee
from (\ref{dsmax}) and (\ref{entapc}) the entanglement between the
halves is
\be
E(\rho)\sim\f{1}{2}\log n,
\ee
and the entropy of one-half of the horizon state is
\be
S(\rho_{n/2})\sim\f{n}{2}\log(1+2k)-\log n.
\ee
As a result,
\be
I_{\rho^s}(\mbox{$\f n2$}:\mbox{$\f n2$})\sim \f{3}{2}\log n.
\ee
\section{Overview of the representations of the permutation group}

We list some of the basic properties of the representations of
$S_{n}$. An exhaustive discussion can be found, e.g., in
\cite{good,che,wkt}. Irreducible representations of a permutation
group are labeled by the Young tableaux, and those that
correspond to the irreps $\sigma_{n,j}$ consist of just two rows.
\be
\begin{tabular}{|c|c|c|c|c|}\hline
1 & 2  &\ldots &\ldots & $n+j$ \\ \hline

$n+j+1$ &\ldots  & $2n$  \\ \cline{1-3}
\end{tabular} \label{Young}
\ee
The Young tableau of Eq.~(\ref{Young}) is a  \emph{normal table},
where the numbers $1,2,\ldots, 2n$ appear in order from left to
right and from the upper row to the lower row. The general
formulas that define the dimensionality of the irreducible
representations of the permutation group
 reduce in the case of $\sigma_{n,j}$ to
$d_j^{(2n)}$ whose properties are described in Appendix C. The
dimension of a representation equals to the number of  distinct
\emph{standard tableaux}, where number in each row appear increasing (not necessarily in strict order)
to the right and in each column appear increasing to the bottom.
The rule for their ordering will be given below. For each
partition $[\nu]$ we enumerate the standard tableaux  as
$([\nu],m)$, where $m=1,\ldots,\dim
\sigma_{[\nu]}$.

 The standard orthonormal
basis for the irreducible representation of $S_n$ that corresponds
to the partition $[\nu]$ (Young-Yamanouchi basis) is labeled by
the irreducible representations of the groups in the subgroup
chain $S_n\supset S_{n-1}\supset\ldots\supset S_2$ to which it
belongs. A simple rule describes this construction: from a given
Young tableau $([\nu],m)$ of $n$ objects one obtains another
tableau $([\nu'],m')$  of $n-1$ objects  by removing the box that
contains the number $n$, etc. For example, a particular basis
vector of the $[31]$ irrep of $S_4$ is labeled by the following
chain of irreps:
\be
\begin{tabular}{|c|c|c|}\hline
1 & 3  &4 \\ \hline

2    \\ \cline{1-1}
\end{tabular} \rightarrow
\begin{tabular}{|c|c|}\hline
1 & 3   \\ \hline

2    \\ \cline{1-1}
\end{tabular} \rightarrow
\begin{tabular}{|c|}\hline
1    \\ \hline

2    \\ \cline{1-1}
\end{tabular} ~,\label{Youngstan}
\ee
i.e. this irreducible basis vector belongs the irrep $[31]$  of
$S_4$, $[21]$ of $S_3$, and [11] of $S_2$. This labeling uniquely
specifies the basis vectors. They are ordered by the
\emph{Yamanouchi symbols} $(r_nr_{n-1}\ldots r_1)$, where $r_i$ is
the row number of $i$. For example, the above vector corresponds
to the Yamanouchi symbol $(1121)$. Once all the standard tableaux
of  $[\nu]$ are labelled by their Yamanouchi symbols, they can be
linearly ordered, with the vector with the largest Yamanouchi
symbol being labelled $|[\nu],1\9$, the vector with the second
largest symbol being $|[\nu],2\9$, etc. In practice, the
irreducible vectors are found as the simultaneous eigenvectors of
the complete commuting set of operators. In the case of a
permutation group, this set consists of the 2-cycle class
operators for the all subgroups of the chain $S_n\supset
S_{n-1}\supset\ldots\supset S_2$.

The matrix elements of all operators that act on the irreducible
representation space $\sigma_{[\nu]}$ of $S_n$ can be
reconstructed  from the representations of $n-1$ generators of
$S_n$, which are the 2-cycles permutations (12),(23),\ldots,
$(n-1,n)$. Their matrix elements in the $|[\nu],m\9$ basis are
obtained from the following set of rules:

1. If $i-1$ and $i$ are in the same row or column of a standard
tableau $([\nu],m)$, then
\be
(i-1,i)|[\nu],m\9=\pm |[\nu],m\9,
\ee
with the plus sign taken when both numbers are in the same row.

2. If $i-1$ and $i$ are not in the same row or column, then the
matrix element $D^{[\nu]}_{m'm}$ is defined with the help of the
\emph{axial distance} $l$. It is calculated as follows: starting
from the box that contains $i-1$ one proceeds by a rectangular
route, stepping one box each time, until  the box with $i$ is
reached. Each step right or upward contributes $+1$ to the axial
distance, while a step downwards or left decreases $s$ by 1. The
matrix elements
\be
D^{[\nu]}_{m'm}\equiv\6[\nu],m'|(i-1,i)|[\nu],m\9
\ee
are given by
\be
D^{[\nu]}_{m'm}=\left\{
\begin{array}{ll}
   1/l,& m'=m \\
   \sqrt{l^2-1}/l,& ([\nu],m')=(i-1,1)([\nu],m) \\
   0 & {\rm otherwise}
\end{array}
\right.
\ee
For a given standard tableau $(r_nr_{n-1}\ldots r_1)$ and a
permutation $(i-1,i)$ the axial distance is
\be
l=c_i-c_{i-1}-(r_i-r_{i-1}),
\ee
 where $c_i$ is a column number of the
object $i$. For the two-row Young tables the column numbers are
easily obtained from the corresponding Yamanouchi symbol as
\be
c_i=\left\{\begin{array}{ll}
2i-\sum_{j=1}^i r_j, & r_i=1\\
\sum_{j=1}^i r_j-i, & r_i=2
\end{array}
\right.
\ee
For example
\be
\left\langle
\begin{tabular}{|c|c|c|}\hline
 1 & 2  &4 \\ \hline

3    \\ \cline{1-1}
\end{tabular}
\right|(34) \left|
\begin{tabular}{|c|c|c|}\hline
 1 & 2  &4 \\ \hline

3    \\ \cline{1-1}
\end{tabular}\right\rangle=\frac{1}{3},\qquad
\left\langle
\begin{tabular}{|c|c|c|}\hline
 1 & 2  &4 \\ \hline

3    \\ \cline{1-1}
\end{tabular}
\right|(34) \left|
\begin{tabular}{|c|c|c|}\hline
 1 & 2  &3 \\ \hline

4    \\ \cline{1-1}
\end{tabular}\right\rangle=\sqrt{\frac{8}{3}}.
\ee

An alternative method of constructing irreps of $S_n$ is to work
on its group algebra $\C S_n$. For a given Young tableau
$\lambda\equiv([\nu],m)$ one defines a subgroup $P\subset S_n$
that preserves the rows of $\lambda$, and the subgroup $Q_\lambda$
that preserves its columns. Then the symmetrizer $a_\lambda$ and
the antysymmetrizer $b_\lambda$ are defined as
\be
a_\lambda=\sum_{g\in P} g, \qquad b_\lambda=\sum_{g\in Q} {\rm
sign}(g) g.
\ee
These two quantities define the irreducible (or Young) symmetrizer
\be
s_\lambda=a_\lambda b_\lambda,
\ee
which generates the irreducible representation $\lambda$ that
corresponds to the partition $[\nu]$ by right multiplication on
$\C S_n$. The symmetrizers that are obtained from different
standard tableaux of the same partition lead to the distinct
equivalent irreps.



\end{document}